\documentclass{elsart}
\usepackage{times,mathptmx}

\usepackage{graphicx}
\usepackage{amssymb}
\usepackage{supertabular}

\begin{document}

\begin{frontmatter}

\title{Measurements of Proton, Helium and Muon Spectra at Small 
Atmospheric Depths with the BESS Spectrometer}
\author[tok]{K. Abe\corauthref{cor1}\thanksref{kobe}},
\corauth[cor1]{Corresponding author.}
\ead{abe@icepp.s.u-tokyo.ac.jp}
\author[tok]{T. Sanuki},
\author[tok]{K. Anraku\thanksref{kanagawa}},
\author[tok]{Y. Asaoka\thanksref{icrr}},
\author[tok]{H. Fuke},
\author[tok]{S. Haino},
\author[kob]{N. Ikeda},
\author[tok]{M. Imori},
\author[tok]{K. Izumi},
\author[kob]{T. Maeno\thanksref{cern}},
\author[kek]{Y. Makida},
\author[tok]{S. Matsuda},
\author[tok]{N. Matsui},
\author[kob]{T. Matsukawa},
\author[tok]{H. Matsumoto},
\author[nas]{J. W. Mitchell},
\author[nas]{A. A. Moiseev},
\author[tok]{J. Nishimura},
\author[kob]{M. Nozaki},
\author[tok]{S. Orito\thanksref{ori}},
\author[nas]{J. F. Ormes},
\author[kek]{M. Sasaki\thanksref{nasa}},
\author[mrl]{E. S. Seo},
\author[kob]{Y. Shikaze},
\author[tok]{T. Sonoda},
\author[nas]{R. E. Streitmatter},
\author[kek]{J. Suzuki},
\author[kek]{K. Tanaka},
\author[kob]{K. Tanizaki},
\author[isa]{T. Yamagami},
\author[kek]{A. Yamamoto},
\author[tok]{Y. Yamamoto},
\author[kob]{K. Yamato},
\author[kek]{T. Yoshida},
\author[kek]{K. Yoshimura}

\address[tok]{The University of Tokyo,
 Bunkyo, Tokyo 113-0033, Japan}
\address[kob]{Kobe University,
 Kobe, Hyogo 657-8501, Japan}
\address[kek]{High Energy Accelerator Research Organization (KEK),
 Tsukuba, Ibaraki 305-0801, Japan}
\address[nas]{National Aeronautics and Space Administration (NASA),
 Goddard Space Flight Center, Greenbelt, MD 20771, USA.}
\address[mrl]{University of Maryland, College Park, MD 20742, USA}
\address[isa]{Institute of Space and Astronomical Science (ISAS),
Sagamihara, 229-8510, Japan}

\thanks[kobe]{Present address: Kobe University,
 Kobe, Hyogo 657-8501, Japan}
\thanks[kanagawa]{Present address: Kanagawa University,
 Yokohama, Kanagawa 221-8686, Japan}
\thanks[icrr]{Present address: ICRR, The University of Tokyo,
 Kashiwa, Chiba 227-8582, Japan}
\thanks[cern]{Present address: CERN, CH-1211 Geneva 23, Switzerland}
\thanks[ori]{deceased.}
\thanks[nasa]{Present address:
 National Aeronautics and Space Administration,
 Goddard Space Flight Center, Greenbelt, MD 20771, USA; 
NAS/NRC Research Associate.}

\begin{abstract}
The cosmic-ray proton, helium, and muon spectra at small 
atmospheric depths of 4.5 -- 28~g/cm$^2$ were precisely measured 
during the slow descending period of the BESS-2001 balloon flight. 
The variation of atmospheric secondary particle fluxes  
as a function of atmospheric depth provides fundamental information to 
study hadronic interactions of the primary cosmic rays with the 
atmosphere.  
\end{abstract}

\end{frontmatter}

\section{Introduction}
\label{sec:intro}
\indent
Primary cosmic rays interact with nuclei in the atmosphere, 
and produce 
atmospheric secondary particles, such as muons, gamma rays and 
neutrinos. 
It is important to understand 
these interactions to investigate cosmic-ray phenomena inside the 
atmosphere. 
For precise study of the atmospheric neutrino 
oscillation~\cite{kamioka}, it is crucial to reduce 
uncertainties in hadronic interactions, which are main sources of 
systematic errors 
in the prediction of the energy spectra of atmospheric neutrinos. 
At small atmospheric depths below a few ten 
g/cm$^2$, production process of muons is predominant over decay process, 
thus we can clearly observe a feature of the hadronic interactions. 
In spite of their importance, 
only a few measurements have been performed with modest statistics 
because of strong constraints of short observation time of a 
few hours 
during balloon ascending periods from the ground to the balloon 
floating altitude. 

In 2001, using the BESS spectrometer, precise measurements of the 
cosmic-ray fluxes and their dependence on atmospheric depth were 
carried out during 
slow descending from 4.5~g/cm$^2$ to 28~g/cm$^2$ for 12.4 hours. 
The growth curves of the cosmic-ray fluxes were precisely measured. 
The results were compared with the predictions based on the 
hadronic interaction models currently used in the atmospheric neutrino 
flux calculations.

\section{BESS spectrometer}
\label{detect}
\begin{figure}[p]
  \includegraphics[width=13cm]{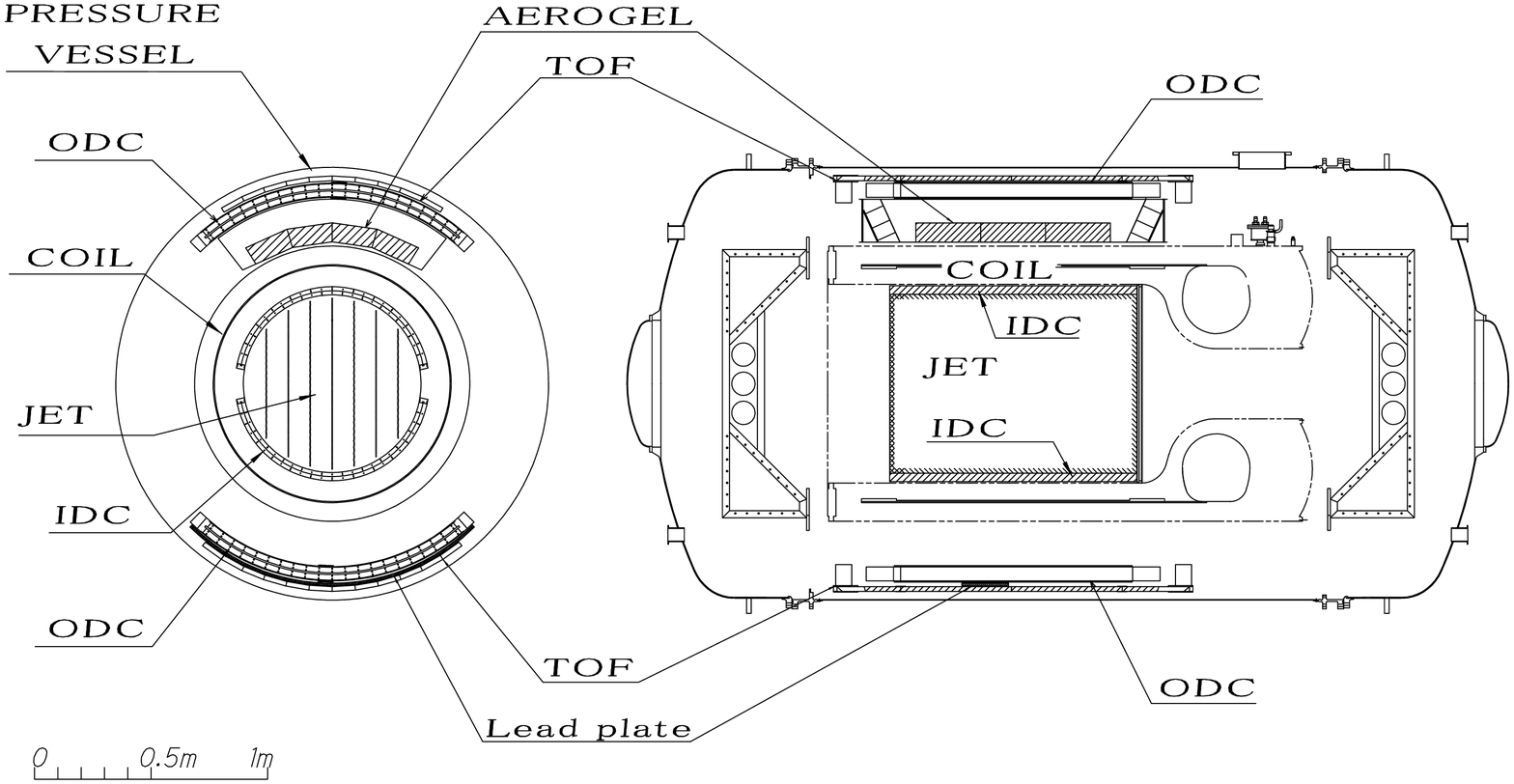}
  \caption{Cross sectional view of the BESS detector.}
  \label{fig:besscross}
\end{figure}

The BESS ($\underline{\rm B}$alloon-borne $\underline{\rm E}$xperiment
with a $\underline{\rm S}$uperconducting 
$\underline{\rm S}$pectrometer)
detector \cite{orito,yamamoto1994,agel,detector,newtof} is
a high-resolution spectrometer with a large acceptance
to perform precise measurement of absolute fluxes
of various cosmic rays~\cite{primary,motoki,norikura},  
as well as highly sensitive searches
for rare cosmic-ray components. 
Fig.~\ref{fig:besscross} shows a schematic cross-sectional view
of the BESS instrument.
In the central region,
a uniform magnetic field of 1 Tesla is provided
by a thin superconducting solenoidal coil. 
The magnetic-rigidity ($R \equiv Pc/Ze$) of
an incoming charged particle is measured
by a tracking system,
which consists of a jet-type drift chamber (JET)
and two inner-drift-chambers (IDC's)
inside the magnetic field. 
The deflection ($R^{-1}$) is calculated 
for each event by applying a circular fitting using up-to 28 hit-points,
each with a spatial resolution of 200~$\mu$m. 
The maximum detectable rigidity (MDR) was estimated to be 200~GV. 
Time-of-flight (TOF) hodoscopes provide the velocity ($\beta$)
and energy loss ($\d E/\d x$) measurements. 
A 1/$\beta$ resolution of 1.4~\% was achieved in this experiment. 
For particle identification,  
the BESS spectrometer was equipped with
a  threshold-type aerogel 
Cherenkov counter and an electromagnetic shower counter. 
The refractive index of silica aerogel radiator was 1.022, and the 
threshold kinetic energy for proton was 3.6~GeV. 
The shower counter consists of a plate of 
lead with two radiation lengths 
covering a quarter area of the lower TOF counters, 
whose output signal was utilized for $e$/$\mu$ identification.

The data acquisition sequence is initiated by a first-level TOF trigger, 
which is a simple coincidence of signals in the upper and lower TOF 
counter.  In order to build a sample of unbiased triggers, one of every 
four events were recorded. 
The TOF trigger efficiency was evaluated to be 99.4~\% $\pm$ 0.2~\% by using 
secondary proton beam at the KEK 12~GeV proton synchrotron.
In addition to the TOF trigger, an auxiliary trigger is generated by a 
signal from the Cherenkov counter to record  particles above 
threshold energy without bias or sampling.  
An efficiency of Cherenkov trigger were evaluated as a ratio 
of the Cherenkov-triggered events among the unbiased trigger sample. 
It was 95.1~\% for relativistic particles ($\beta\rightarrow1$).
For flux determinations, 
the Cherenkov-triggered events were used only  
above 9.5~GV for protons, 11.1~GV for helium nuclei,  
and 0.90~GV for muons. 
Below these rigidities, the TOF-triggered events were used.

\section{Balloon flight}
\label{sec:data}

\begin{figure}[p]
  \begin{center}
    \includegraphics[width=13cm]{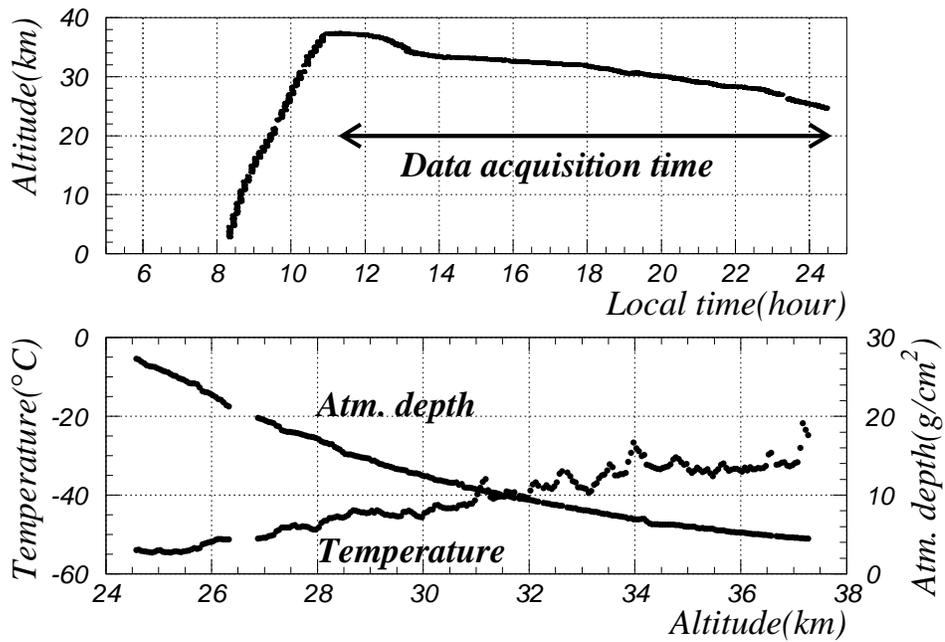}
  \end{center}
  \vspace{-0.5cm}
  \caption{Altitude during the BESS-2001 balloon flight 
    experiment(top).    
    Temperature and residual atmosphere as a function of 
    altitude(bottom).}
  \label{fig:monitor}
\end{figure}

The BESS-2001 balloon flight was carried out at Ft. Sumner, New Mexico, 
USA (34$^{\circ}$49$'$N, 104$^{\circ}$22$'$W) on 24th September 2001. 
Throughout the flight, the vertical geomagnetic cut-off rigidity 
was about 4.2~GV.
The balloon reached at a normal floating altitude of 36~km at an 
atmospheric depth of 4.5~g/cm$^2$. After a few hours, the balloon 
started to lose the floating altitude and
continued descending for more than 13 hours until termination of the 
flight.
During the descending period, data were collected at atmospheric depths
between 4.5 g/cm$^2$ and 28 g/cm$^2$.
The atmospheric depth was measured with accuracy of 
$\pm$ 1~g/cm$^2$, which comes mainly from an error in absolute 
calibration of an environmental monitor system.  
Fig.~\ref{fig:monitor} shows a balloon flight profile   
during the experiment. 

\section{Data analysis}
\label{sec:analysis}

\subsection{Data reduction}
\label{sec:reduction}

We selected ``non-interacted'' events passing through the detector 
without any interactions. 
The non-interacted event was defined as an event, 
which has only one isolated track, 
one or two hit-counters in each layer of the TOF hodoscopes,
and proper d$E$/d$x$ inside the upper TOF counters.  
There is a slight probability that particles 
interact with nuclei in the detector material  
and only one secondary particle goes into the tracking volume to be 
identified as a non-interacted event. 
These events were rejected by requiring proper d$E$/d$x$ inside the upper 
TOF counters, because the interaction in the detector is expected 
to give a large energy deposit in the TOF counter. 
According to the Monte Carlo simulation with GEANT~\cite{geant},  
these events are to be considered only below a few~GeV.
In order to estimate an efficiency of the non-interacted 
event selection, Monte Carlo simulations were performed. 
The probability 
that each particle can be identified as a non-interacted event 
was evaluated
by applying the same selection criterion to the Monte Carlo events as was 
applied to the observed data. 
The systematic error was estimated by 
comparing the hit number distribution of the TOF counters.  
For muons, the simulated data were compared with muon data sample 
measured on the ground.
The systematic error for protons below 1~GeV was directly determined 
by the detector beam test using accelerator proton beam~\cite{beamp}. 
The resultant efficiency and its error of non-interacted event 
selection for protons was 83.3 $\pm$ 2.0~\% at 1.0~GeV 
and 77.4 $\pm$ 2.5~\% at 10~GeV, 
and that for helium nuclei was  
71.6 $\pm$ 2.6~\% at 1.0~GeV/n and 66.0 $\pm$ 2.9~\% at 10~GeV/n. 
The efficiency for muons was 
94.0 $\pm$ 0.9~\% and 92.8 $\pm$ 
0.8~\% at 0.5~GeV and 10~GeV, respectively. 

The selected non-interacted events were required to pass through the  
fiducial volume defined in this analysis. 
The fiducial volume of the detector 
was limited to the central region of the JET 
chamber 
for a better rigidity measurement. 
The zenith angle ($\theta _z$) was limited 
within $\cos \theta _z \geq 0.90$ to obtain nearly vertical fluxes.  
For the muon analysis, we used only the particles which passed through 
the lead plate to estimate electron contaminations.
For the proton and helium analysis, 
particles were required not to pass through the lead plate 
so as to keep the interaction probability inside the detector 
as low as possible.

In order to check the track reconstruction efficiency inside the 
tracking system, 
the recorded events were scanned randomly.  
It was confirmed that  
996 out of 1,000 visually identified tracks which passed through
the fiducial volume were successfully reconstructed, 
thus the track reconstruction efficiency was evaluated to be 
99.6 $\pm$ 0.2~\%. 
It was also confirmed that rare interacted events are fully 
eliminated by the non-interacted event selection.

\subsection{Particle identification}
\label{sec:reduction}

\begin{figure}[pht!]
  \begin{center}
    \includegraphics[width=13cm]{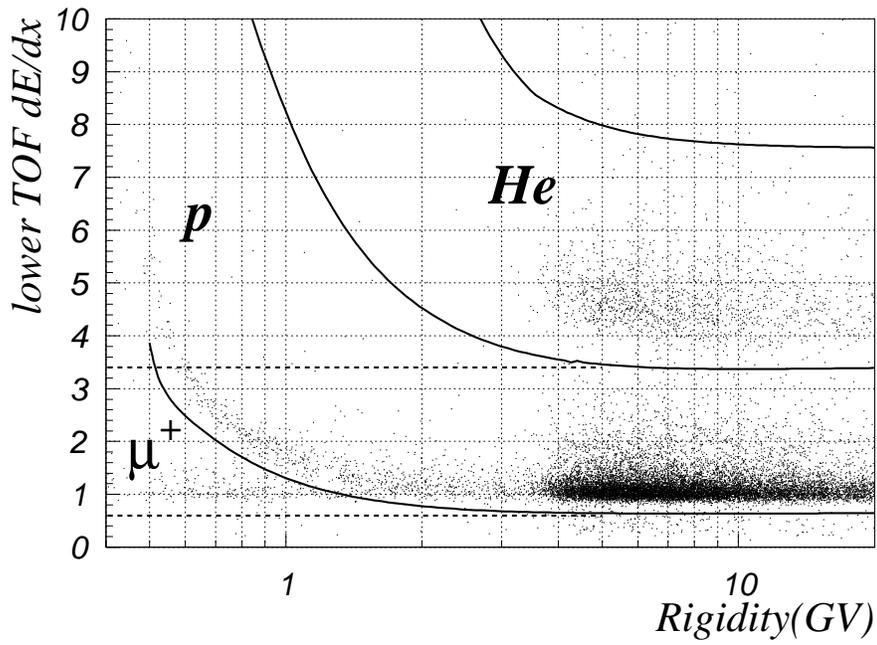}
  \end{center}
  \caption{Distribution of d$E$/d$x$ inside the lower TOF counters. 
	Solid lines 
	show selection
      boundaries for protons and helium nuclei. Dashed lines show those
      for muons.}
  \label{fig:pidde}
\end{figure}

In order to select singly and doubly charged particles, 
particles were required to have proper d$E$/d$x$
as a function of rigidity 
inside both the upper and lower TOF hodoscopes. 
The upper TOF d$E$/d$x$ was already examined 
in the non-interacted event selection.
The distribution of d$E$/d$x$ inside the lower TOF counter and 
the selection boundaries are shown in Fig.~\ref{fig:pidde}. 

In order to estimate efficiencies of the d$E$/d$x$ selections for protons 
and helium nuclei, we used another data sample selected 
by independent information of energy loss inside the JET chamber. 
The estimated efficiency in the d$E$/d$x$ selection at 1~GeV was 98.3 
$\pm$ 0.4~\% and 97.2 $\pm$ 0.5~\% 
for protons and helium nuclei, respectively.
The accuracy of the efficiency for protons and helium nuclei was limited 
by statistics of the sample events.
Since muons could not be distinguished from electrons by the JET chamber, 
the d$E$/d$x$ selection efficiency for muons was estimated by the Monte
Carlo simulation to be 99.3 $\pm$ 1.0~\%. 
The error for muons comes from the discrepancy between the observed 
and simulated d$E$/d$x$ distribution inside the TOF counters.
The d$E$/d$x$ selection efficiencies were almost constant 
in the whole energy region discussed here.  

\begin{figure}[pht!]
  \begin{center}
    \includegraphics[width=13cm]{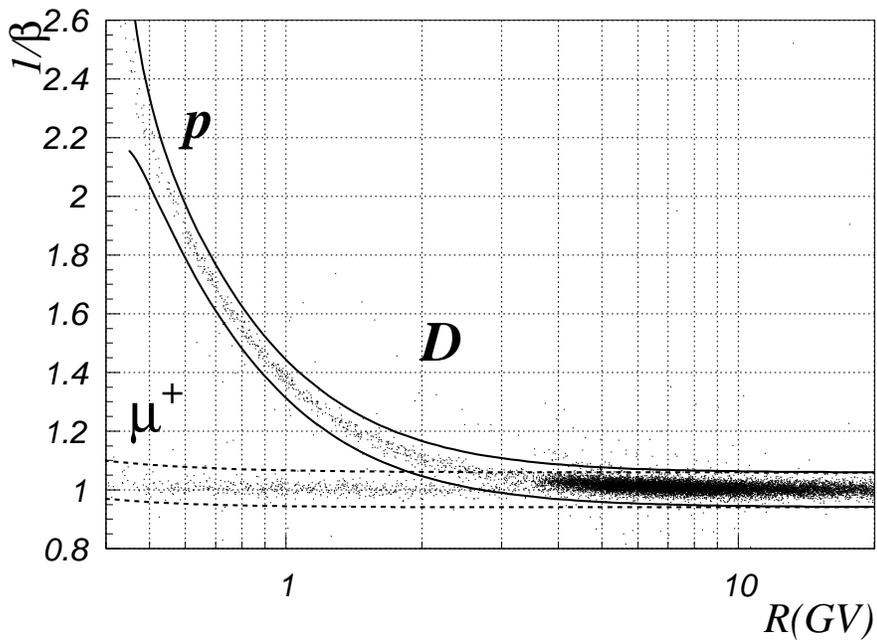}
  \end{center}
  \caption{Distribution of 1/$\beta$ for $Z$=1 particles. 
	Solid lines show the selection
	boundaries for protons.
	Dashed lines show those for muons.}
  \label{fig:pidbt}
\end{figure}

Particle mass was reconstructed by using the relation of 1/$\beta$, 
rigidity and charge, 
and was required to be consistent with protons, helium nuclei or muons.
An appropriate relation between 1/$\beta$ 
and rigidity for each particle  
was required as shown in Fig.~\ref{fig:pidbt}. 
Since the 1/$\beta$ distribution is well described by Gaussian and a 
half-width of the 1/$\beta$ selection band was set at 3.89 $\sigma$, 
the efficiency is very close to unity (99.99~\% for pure Gaussian). 

\subsection{Contamination estimation}
\label{contami}
\subsubsection{Protons}
Protons were clearly identified without any contamination 
below 1.7~GV by the mass selection, as shown in Fig.~\ref{fig:pidbt}.
Above 1.7~GV, however, light 
particles such as positrons and muons contaminate proton's 
$1/\beta$-band, and  above 4~GV deuterons ($D$'s) start to contaminate 
it. 

To distinguish protons from muons and positrons above 1.7~GV,  
we required that light output of the aerogel Cherenkov counter 
should be smaller than the threshold.
This requirement rejected 96.5~\% of muons and positrons
while keeping the efficiency for protons as high as 99.5~\%.
The aerogel Cherenkov cut was applied below 3.7~GV, 
above which Cherenkov output for protons begins to increase rapidly. 

Contamination of muons in proton candidates after the aerogel Cherenkov 
cut was estimated and subtracted 
by using calculated muon fluxes~\cite{HONDA}, 
which was normalized to the observed fluxes below 1.0~GeV 
where positive muons were clearly separated from protons.
The positron contamination was calculated from the normalized 
positive muon fluxes and observed $e$/$\mu$ ratios.  
The resultant positive muon and positron contamination in the proton 
candidates was less than 0.8~\% and 4.1~\% below and above 3.7~GV, 
respectively, at 26.4~g/cm$^2$ where observed $(\mu$+$e)$/$p$ ratio is 
largest during the experiment. 
The error in this subtraction was estimated from the ambiguity in the 
normalization of the muon spectra to be less than 0.2~\% and 0.8~\% 
below and above 3.7~GV, respectively.

Since the geomagnetic cut-off rigidity is 4.2~GV, 
most of deuterons contaminating proton candidates above 4~GV
are considered to be primary cosmic-ray particles.
The primary $D$/$p$ ratio was estimated by our previous 
measurement carried out in 1998 -- 2000 at Lynn Lake, 
Manitoba, Canada, 
where the cut-off rigidity is as low as 0.5~GV.
The $D$/$p$ ratio was found to be 2~\% at 3~GV.
No subtraction was made for deuteron contamination
because there is no reliable measurement of the deuteron 
flux above 4~GV.
Therefore, above 4~GV hydrogen nuclei selected which 
included a small amount of deuterons.
The $D$/$p$ ratio at higher energy is expected to decrease~\cite{seo}
due to
the decrease in escape path lengths of primary cosmic-ray 
nuclei~\cite{engelm} and
the deuteron component is as small as the statistical error 
of the proton flux.

\subsubsection{Helium nuclei}
Helium nuclei were clearly identified by using both upper and lower 
TOF d$E$/d$x$ as shown in Fig.~\ref{fig:pidde}. 
Landau tail in proton's d$E$/d$x$ might 
contaminate the helium d$E$/d$x$ band, however 
this contamination from protons was as small as 
3$\times$10$^{-4}$. It was estimated 
by using another sample of proton data selected 
by d$E$/d$x$ in the JET chamber. 
No background subtraction was made for helium.
Obtained helium fluxes include both $^3$He and $^4$He.  
 
\subsubsection{Muons}
\begin{figure}[pht!]
  \begin{center}
    \includegraphics[width=13cm]{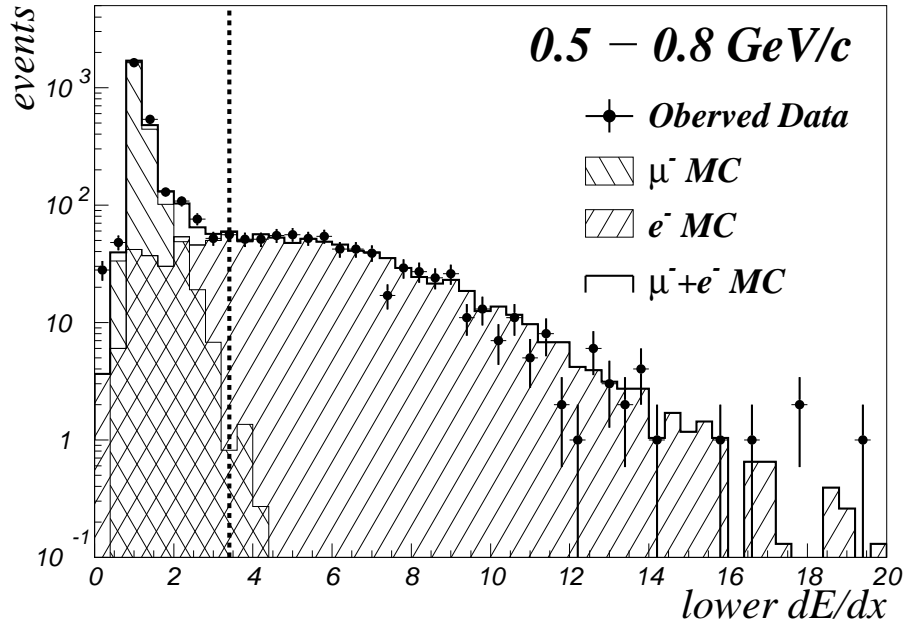}
  \end{center}
  \caption{Observed and simulated distribution of d$E$/d$x$ 
      inside the lower TOF counters for the events in which 
      track passed through the lead plate.
      Dashed line shows the selection boundary for muon.
      The observed data were summed up irrespective of the 
      residual atmosphere.}
  \label{fig:ebackg}
\end{figure}

Electrons and pions could contaminate muon candidates. 
To estimate electron contamination,  
we used the  d$E$/d$x$ information inside 
the lower TOF counters covered with the lead plate.  
We calculated lower TOF d$E$/d$x$ distribution for muons and electrons 
using the Monte Carlo simulation. 
The most adequate e/$\mu$ ratio was estimated by changing weights 
both for muons and electrons so as to reproduce the observed d$E$/d$x$ 
distribution. 
The simulated distributions well-agreed with the observed data 
as shown in Fig.~\ref{fig:ebackg}.
The electron contamination was about 10~\% of muon candidates  
at 0.5~GV, and less than 1\% above 1~GV.
Since in the lower energy region, 
the difference between the d$E$/d$x$ 
distributions for electrons and muons are small, 
rejection power of d$E$/d$x$ selection are lower.
The error 
was estimated during the fitting procedure to be about 5~\% at 0.5~GV 
and 0.1~\% at 8.0~GV.
Accuracy of the electron subtraction was limited by poor statistics 
of electron events. 
For the pion contamination we made no subtraction, thus 
the observed muon fluxes include pions. 
According to a theoretical calculation~\cite{steph}, 
the $\pi$/$\mu$ ratio at 
a residual atmosphere of 3~g/cm$^2$ through 10~g/cm$^2$ 
was less than 3~\% at 1~GV and less than 10~\% at 10~GV.

\begin{figure}[pht!]
  \begin{center}
    \includegraphics[width=13cm]{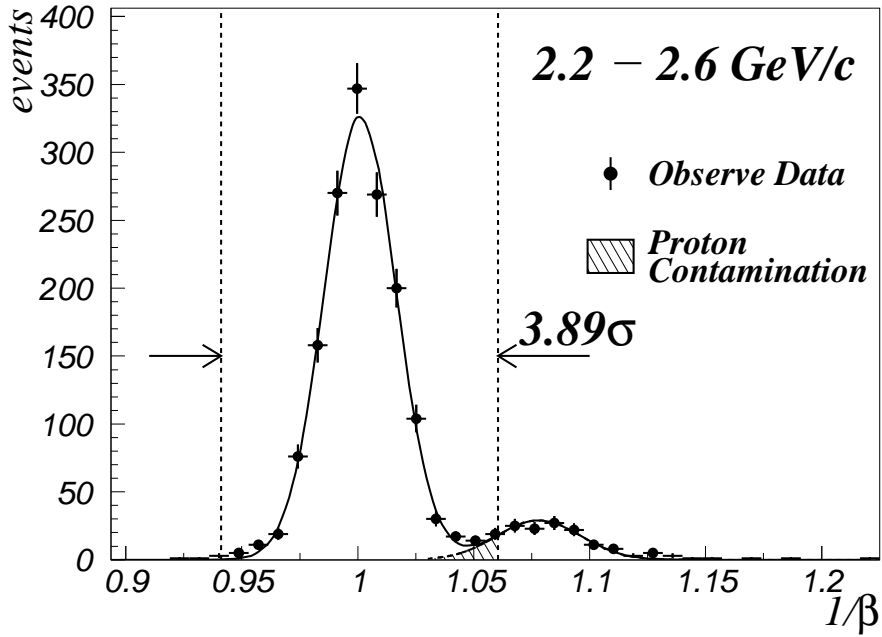}
  \end{center}
  \caption{An estimation of proton contamination by the TOF information.
    Hatched area shows an estimated proton contamination.}
  \label{fig:pbackg}
\end{figure}

Between 1.6 and 2.6~GV, protons contamination in the positive 
muon candidates 
and its error were estimated by fitting the 1/$\beta$ distribution 
with a double-Gaussian function as shown in Fig~\ref{fig:pbackg}. 
The estimated ratio of proton contamination in muon candidates 
was less than 3.3~$\pm$~1.3~\%.


\subsection{Corrections}
\label{correct}
In order to determine the proton, helium and muon spectra, 
ionization energy loss inside the detector material, live-time  
and geometrical acceptance need to be estimated.  

The energy of each particle at the top of the instrument was calculated 
by summing up the ionization energy losses inside the instrument 
with tracing back the event trajectory. 
The total live data-taking time was measured exactly 
to be 40,601 seconds by counting 1 MHz clock 
pulses with a scaler system gated by a ``ready'' status that control the 
first level trigger.
The geometrical acceptance defined for this analysis was calculated 
as a function of rigidity by using simulation 
technique~\cite{sullivan1971}. 
In the high rigidity region where a track of the particle is nearly 
straight, 
the geometrical acceptance is 0.097 ${\mathrm {m^2sr}}$ for protons and 
helium nuclei, 
and 0.030 ${\mathrm {m^2sr}}$ for muons. 
The acceptance for muons is about 1/3 of that for protons and helium 
nuclei, because we required muons to pass through the 
lead plate while protons and helium nuclei were required not to pass 
through the lead plate.
The simple cylindrical shape and the uniform magnetic field
 make it simple and reliable
 to determine the precise geometrical acceptance. 
The error which arose from uncertainty of the detector alignment
 was estimated to be 1~\%. 

\section{Results and discussions}
\begin{figure}[pht!]
  \begin{center}
    \includegraphics[width=13cm]{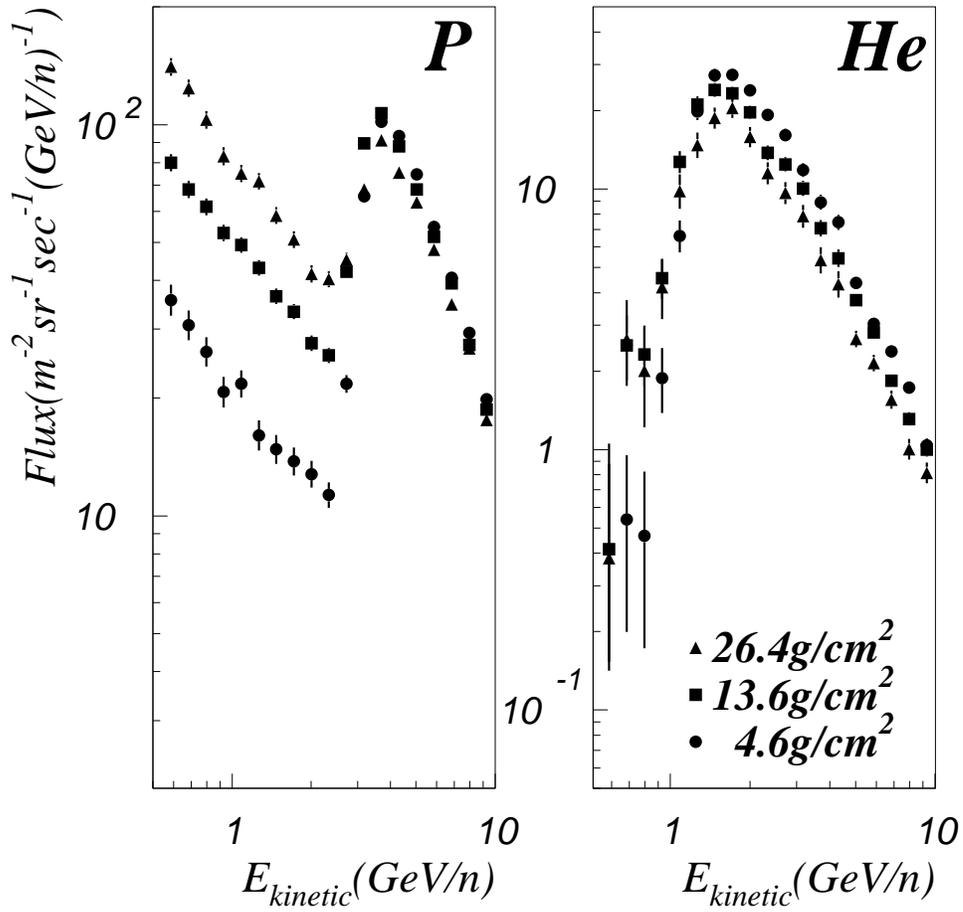}
  \end{center}
  \caption{The observed proton and helium spectra. 
	The error bars include statistical error only.
	In the proton spectra, atmospheric secondary components are 
	clearly observed below 2.5~GeV in the BESS results.}
  \label{fig:prevphe}
\end{figure}

\begin{figure}[pht!]
  \begin{center}
    \includegraphics[width=13cm]{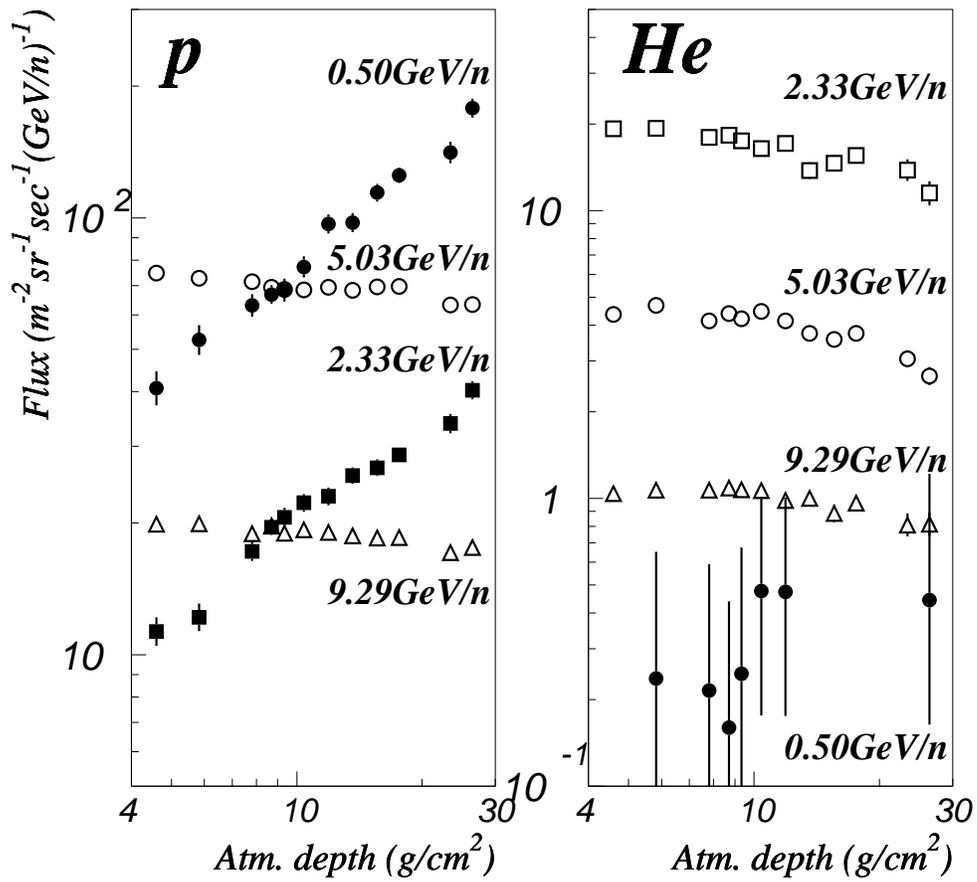}
  \end{center}
  \caption{The observed proton and helium fluxes 
	as a function of atmospheric depth.}
  \label{fig:phegrowth}
\end{figure}

The proton and helium fluxes in energy ranges of 0.5--10~GeV/n
and muon flux in 0.5~GeV/$c$--10~GeV/$c$, at small atmospheric depths 
of 4.5~g/cm$^2$ through 28~g/cm$^2$, 
have been obtained from the BESS-2001 balloon flight. 
The results are summarized in Table~\ref{tab:sump1}. 
The statistical errors were calculated as 68.7~\% confidence interval   
based on Feldman and Cousins's ``unified approach'' \cite{Feld98}.
The overall errors including both statistic and systematic 
errors are less than 8~\%, 10\% and 20~\% 
for protons, helium nuclei and muons, respectively. 
The obtained proton and helium spectra are 
shown in Fig.~\ref{fig:prevphe}. 
Around at 3.4~GeV for protons and 1.4~GeV/n for helium nuclei, 
a geomagnetic cut-off effect is clearly observed in their spectra. 
The proton spectrum measured by the AMS experiment in 1998~\cite{AMSre} 
at the similar geomagnetic 
latitude ($0.7<\Theta_{M}<0.8$, where $\Theta_{M}$ is 
the corrected geomagnetic latitude~\cite{mlat}) is also shown in 
Fig.~\ref{fig:prevphe}.
The AMS measured proton spectra in space (at an altitude of 380~km), 
which are free from atmospheric secondary particles\footnote{The AMS 
observed substantial ``second'' spectra below the geomagnetic cut-off. 
Most of them follow a complicated trajectory in the Earth's magnetic 
field, and could not be observed at balloon altitude.}.  
In the BESS results 
the atmospheric secondary spectra for protons below 2.5~GeV 
are observed.  
Fig.~\ref{fig:phegrowth} shows the observed proton 
 and helium fluxes as a function of the atmospheric depth.
Below 2.5~GeV the proton fluxes clearly increase as the atmospheric depth 
increases. 
It is because the secondary protons are produced in the atmosphere. 
In the primary fluxes above the geomagnetic cut-off,  
the fluxes attenuate as the atmospheric depth increases.  
In this energy region, the production of the secondary protons is  
much smaller 
than interaction loss of the primary protons.  
This is because the flux of parent particles of secondary protons is much
smaller due to the steep spectrum of primary cosmic rays.  

\begin{figure}[pht!]
  \begin{center}
    \includegraphics[width=13cm]{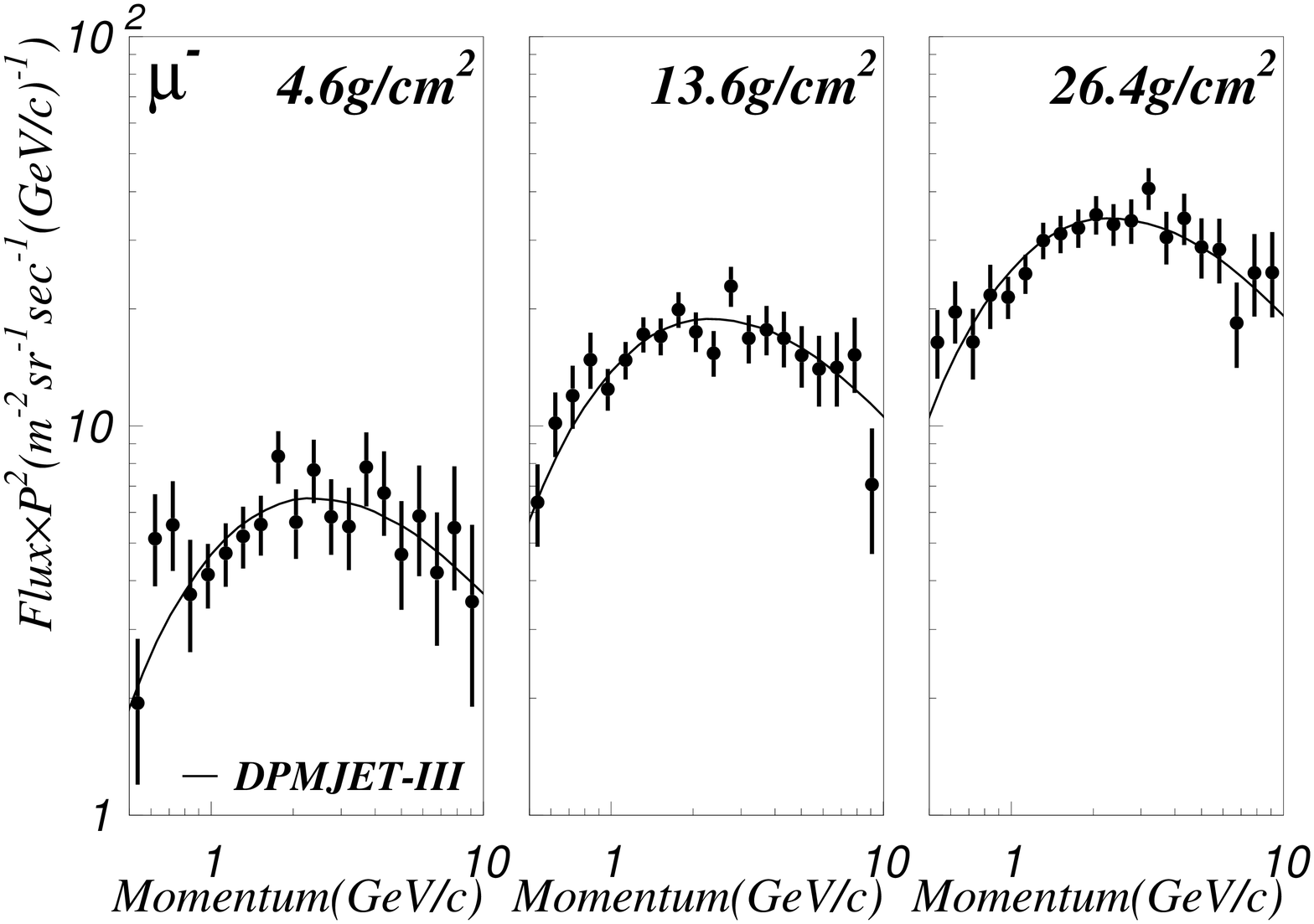}
  \end{center}
  \caption{The observed negative muon spectra.
	The solid lines show theoretical predictions calculated by 
	using DPMJET-III.}
  \label{fig:diffintm}
  \begin{center}
    \includegraphics[width=13cm]{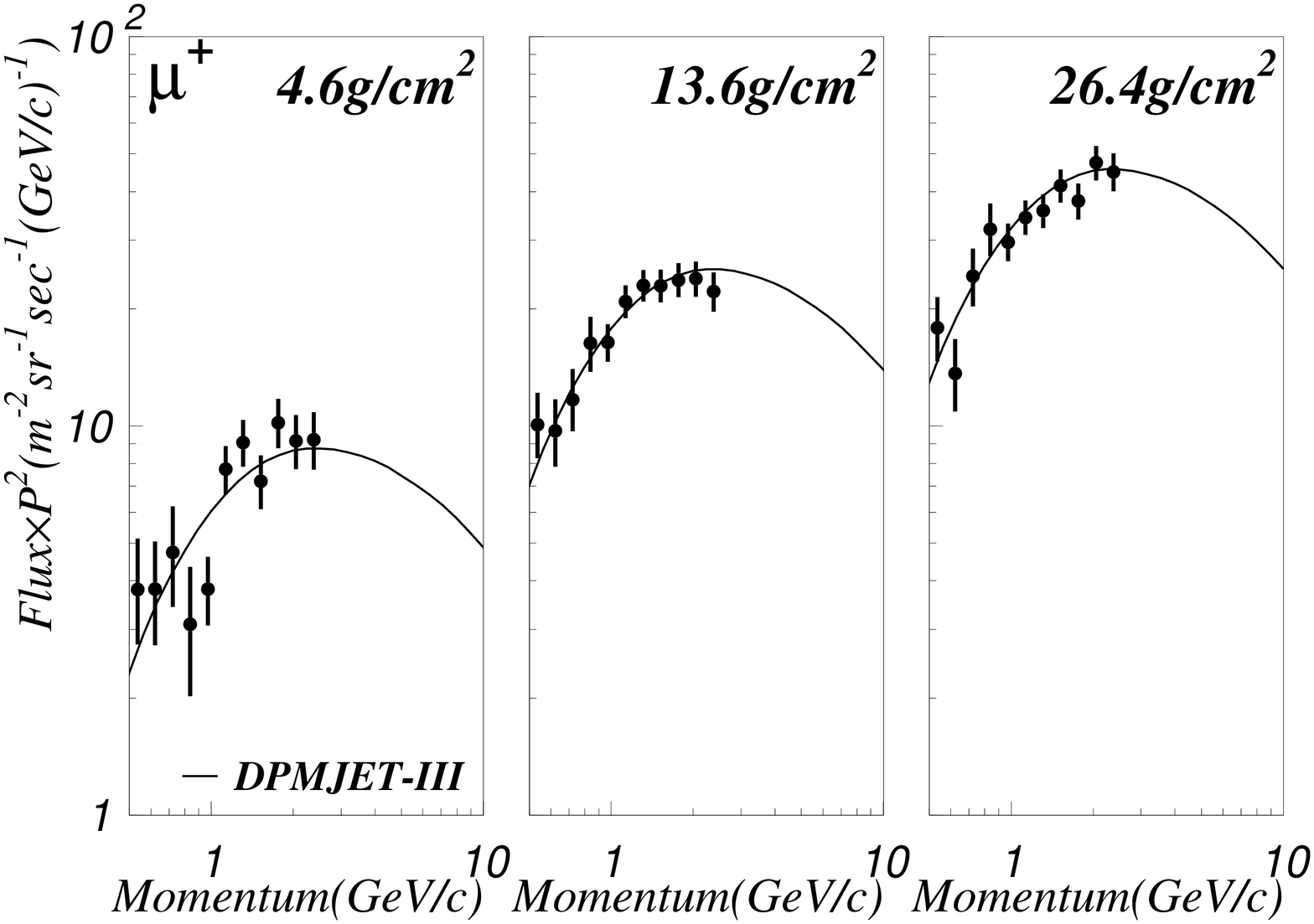}
  \end{center}
  \caption{The observed positive muon spectra.
	The error bars include statistical error only.
	The solid lines show theoretical predictions calculated by 
	using DPMJET-III.}
  \label{fig:diffintp}
\end{figure}

\begin{figure}[pht!]
  \begin{center}
    \includegraphics[width=13cm]{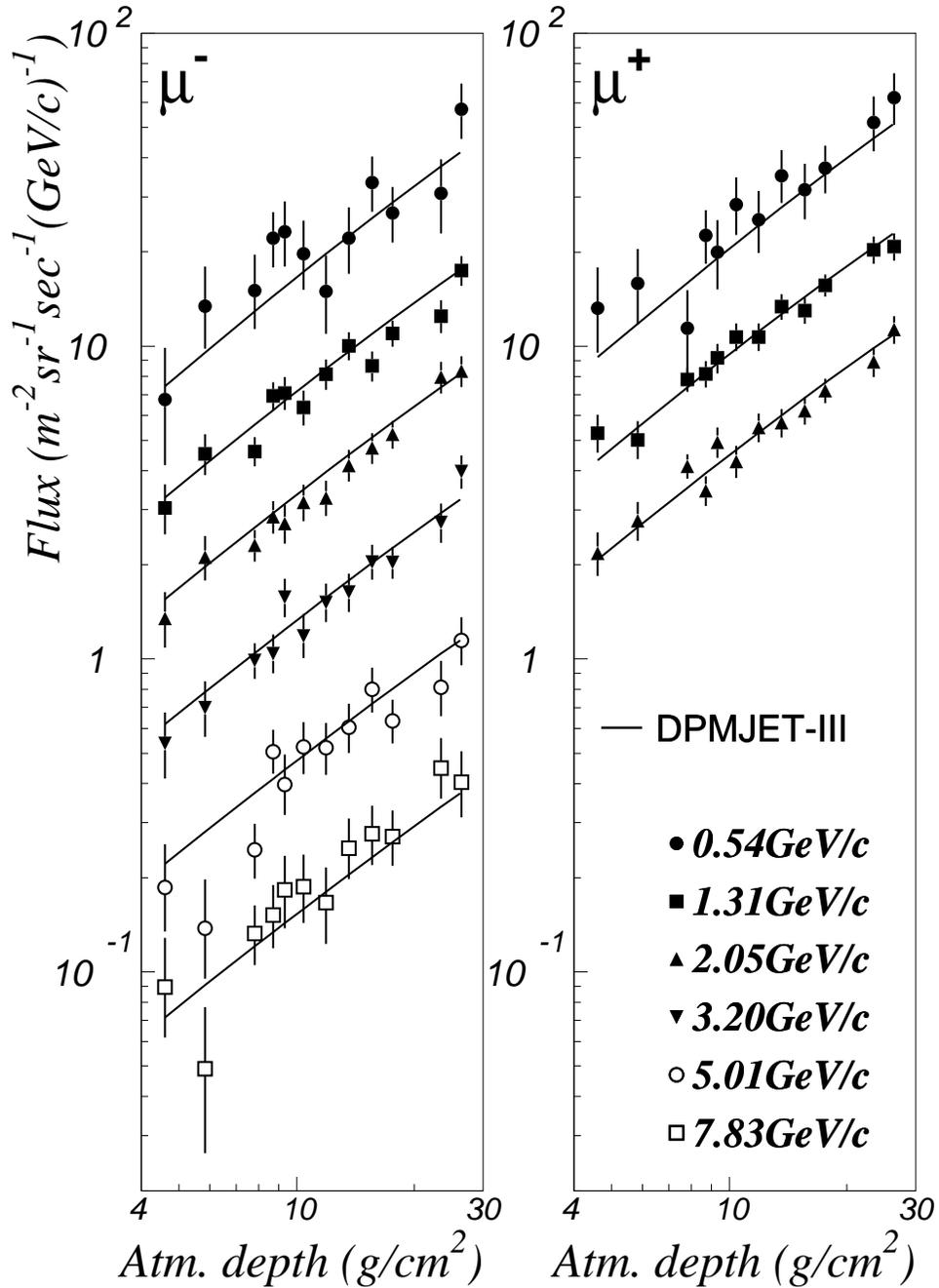}
  \end{center}
  \caption{The observed negative and positive muon fluxes. 
	The solid lines show theoretical predictions calculated by 
	using DPMJET-III.}
  \label{fig:mgrowth}
\end{figure}

Figs.~\ref{fig:diffintm} and \ref{fig:diffintp} show the observed 
muon spectra together with theoretical predictions. 
The predictions were made with the hadronic interaction model, 
DPMJET-III~\cite{DPMJET}, which was 
used for the evaluation of atmospheric neutrino 
fluxes~\cite{HONDA}.
The obtained proton fluxes were used 
to reproduce the primary cosmic-ray fluxes in the calculation. 
Fig.~\ref{fig:mgrowth} shows the observed muon fluxes as a function of 
atmospheric depth together with the calculated fluxes.  
The calculated fluxes show good agreement with the observed data. 
Further detailed study of the hadronic interaction models will 
be discussed elsewhere. 

\section{Conclusion}
We made precise measurements of cosmic-ray spectra of protons, 
helium nuclei 
and muons at small atmospheric depths of 4.5 through 28~g/cm$^2$, during 
a slow descending period of 12.4 hours, in the BESS-2001 balloon flight 
at Ft. Sumner, New Mexico, USA.  
We obtained the proton and helium fluxes with overall errors of 
8~\% and 10~\%, respectively, in an energy region of 0.5 -- 10~GeV/n. 
The muon fluxes were obtained with an overall error of 20~\% in a 
momentum region of 0.5 -- 10~GeV/$c$.
The results provide fundamental information to investigate 
hadronic interactions of cosmic rays with atmospheric nuclei. 
The measured muon spectra showed good agreement 
with the calculations by 
using the DPMJET-III hadronic interaction model. 
The understanding of the interactions 
will improve the accuracy of calculation of atmospheric neutrino fluxes.

\begin{ack}
We would like to thank NASA and 
the National Scientific Balloon Facility (NSBF) for their
professional and skillful work in carrying out
the BESS flights. 
We are indebted to M. Honda and T. Kajita of ICRR, 
the University of Tokyo  
for their kindest cooperation for Monte Carlo simulations and theoretical 
interpretations. 
We would like to thank ISAS and KEK for their continuous support 
and encouragement for the BESS experiment. 
The analysis was performed with the computing facilities 
at ICEPP, the University of Tokyo. 
This experiment was supported by Grants-in-Aid (12047206 and 12047227)
from the Ministry of Education, Culture, Sport, Science and Technology, 
MEXT. 
\end{ack}

\clearpage



\clearpage

\begin{table}[h!]
\tiny{
  \caption
{Observed proton fluxes}
  \label{tab:sump1}
\renewcommand{\arraystretch}{1.5}
  \begin{center}
\rotatebox{90}{
    \begin{tabular}{cllllllllllll}
      \hline
      \hline
      \begin{tabular}{@{}c@{}}       
      Energy range\\
      (GeV)
      \end{tabular}
      & 
      \multicolumn{12}{c}{
      \begin{tabular}{@{}c@{}}       
	Flux$\pm~\Delta$Flux$_{\rm sta}
	\pm~\Delta$Flux$_{\rm sys}$\\
	(m$^{-2}$sr$^{-1}$s$^{-1}$GeV$^{-1}$)
      \end{tabular}
      }
      \\
      \hline
      &       \multicolumn{12}{c}{atmospheric depth range [mean] (g/cm$^2$)}\\
      &  \multicolumn{1}{c}{4.46--4.80}  
      &  \multicolumn{1}{c}{4.86--7.21}  
      &  \multicolumn{1}{c}{7.04--8.23}  
      &  \multicolumn{1}{c}{8.24--9.08}  
      &  \multicolumn{1}{c}{9.06--9.54}  
      &  \multicolumn{1}{c}{9.60--11.4} 
      &  \multicolumn{1}{c}{11.4--12.5}
      &  \multicolumn{1}{c}{12.7--14.6}  
      &  \multicolumn{1}{c}{14.7--16.4} 
      &  \multicolumn{1}{c}{16.5--19.8}   
      &  \multicolumn{1}{c}{21.2--25.0}   
      &  \multicolumn{1}{c}{25.0--28.2}   \\
      & \multicolumn{1}{c}{[4.58]} & \multicolumn{1}{c}{[5.82]} 
      & \multicolumn{1}{c}{[7.81]} & \multicolumn{1}{c}{[8.70]} 
      & \multicolumn{1}{c}{[9.33]} & \multicolumn{1}{c}{[10.4]} 
      & \multicolumn{1}{c}{[11.9]} & \multicolumn{1}{c}{[13.6]} 
      & \multicolumn{1}{c}{[15.6]}& \multicolumn{1}{c}{[17.6]}
      & \multicolumn{1}{c}{[23.4]}& \multicolumn{1}{c}{[26.4]}\\
      \hline 
    0.46--    0.54
&  40.8$^{+   3.3+   0.9}_{-   3.1-   0.9}$
&  52.5$^{+   3.7+   1.2}_{-   3.5-   1.2}$
&  63.0$^{+   3.8+   1.4}_{-   3.6-   1.4}$
&  66.7$^{+   3.4+   1.5}_{-   3.2-   1.5}$
&  68.3$^{+   4.3+   1.6}_{-   4.0-   1.6}$
&  77.2$^{+   4.5+   1.8}_{-   4.2-   1.8}$
&  96.7$^{+   4.9+   2.2}_{-   4.7-   2.2}$
&  97.5$^{+   5.0+   2.2}_{-   4.8-   2.2}$
&  114.$^{+    5.+    3.}_{-    5.-    3.}$
&  125.$^{+    5.+    3.}_{-    5.-    3.}$
&  141.$^{+    8.+    3.}_{-    8.-    3.}$
&  178.$^{+    9.+    4.}_{-    9.-    4.}$
 \\
    0.54--    0.63
&  35.6$^{+   2.8+   0.8}_{-   2.6-   0.8}$
&  41.0$^{+   3.0+   0.9}_{-   2.8-   0.9}$
&  59.1$^{+   3.4+   1.4}_{-   3.2-   1.4}$
&  57.1$^{+   2.9+   1.3}_{-   2.7-   1.3}$
&  64.4$^{+   3.8+   1.5}_{-   3.6-   1.5}$
&  69.7$^{+   3.9+   1.6}_{-   3.7-   1.6}$
&  79.5$^{+   4.1+   1.8}_{-   3.9-   1.8}$
&  79.9$^{+   4.1+   1.8}_{-   3.9-   1.8}$
&  97.5$^{+   4.6+   2.2}_{-   4.4-   2.2}$
&  101.$^{+    4.+    2.}_{-    4.-    2.}$
&  134.$^{+    7.+    3.}_{-    7.-    3.}$
&  141.$^{+    7.+    3.}_{-    7.-    3.}$
 \\
    0.63--    0.74
&  30.7$^{+   2.4+   0.7}_{-   2.2-   0.7}$
&  34.0$^{+   2.5+   0.8}_{-   2.3-   0.8}$
&  46.6$^{+   2.8+   1.1}_{-   2.6-   1.1}$
&  50.7$^{+   2.5+   1.2}_{-   2.4-   1.2}$
&  50.4$^{+   3.1+   1.2}_{-   2.9-   1.2}$
&  58.9$^{+   3.3+   1.3}_{-   3.1-   1.3}$
&  61.6$^{+   3.3+   1.4}_{-   3.2-   1.4}$
&  68.2$^{+   3.5+   1.6}_{-   3.3-   1.6}$
&  80.7$^{+   3.9+   1.8}_{-   3.7-   1.8}$
&  91.2$^{+   3.7+   2.1}_{-   3.6-   2.1}$
&  98.5$^{+   5.7+   2.3}_{-   5.4-   2.3}$
&  124.$^{+    6.+    3.}_{-    6.-    3.}$
 \\
    0.74--    0.86
&  26.3$^{+   2.1+   0.6}_{-   1.9-   0.6}$
&  31.9$^{+   2.2+   0.7}_{-   2.1-   0.7}$
&  40.5$^{+   2.4+   0.9}_{-   2.2-   0.9}$
&  44.1$^{+   2.1+   1.0}_{-   2.0-   1.0}$
&  44.2$^{+   2.7+   1.0}_{-   2.5-   1.0}$
&  52.9$^{+   2.8+   1.2}_{-   2.7-   1.2}$
&  58.7$^{+   3.0+   1.3}_{-   2.8-   1.3}$
&  61.6$^{+   3.1+   1.4}_{-   2.9-   1.4}$
&  69.0$^{+   3.3+   1.6}_{-   3.1-   1.6}$
&  74.3$^{+   3.1+   1.7}_{-   3.0-   1.7}$
&  89.9$^{+   5.0+   2.1}_{-   4.7-   2.1}$
&  103.$^{+    5.+    2.}_{-    5.-    2.}$
 \\
    0.86--    1.00
&  20.8$^{+   1.7+   0.5}_{-   1.6-   0.5}$
&  25.8$^{+   1.9+   0.6}_{-   1.7-   0.6}$
&  36.4$^{+   2.1+   0.8}_{-   2.0-   0.8}$
&  38.9$^{+   1.8+   0.9}_{-   1.7-   0.9}$
&  37.2$^{+   2.2+   0.9}_{-   2.1-   0.9}$
&  46.7$^{+   2.5+   1.1}_{-   2.3-   1.1}$
&  52.4$^{+   2.6+   1.2}_{-   2.5-   1.2}$
&  52.8$^{+   2.6+   1.2}_{-   2.5-   1.2}$
&  59.1$^{+   2.8+   1.4}_{-   2.7-   1.4}$
&  63.4$^{+   2.6+   1.5}_{-   2.5-   1.5}$
&  76.3$^{+   4.3+   1.7}_{-   4.0-   1.7}$
&  83.0$^{+   4.5+   1.9}_{-   4.3-   1.9}$
 \\
    1.00--    1.17
&  21.7$^{+   1.6+   0.4}_{-   1.5-   0.4}$
&  26.4$^{+   1.7+   0.5}_{-   1.6-   0.5}$
&  30.2$^{+   1.7+   0.5}_{-   1.7-   0.5}$
&  36.2$^{+   1.6+   0.6}_{-   1.6-   0.6}$
&  37.5$^{+   2.1+   0.7}_{-   2.0-   0.7}$
&  41.1$^{+   2.1+   0.7}_{-   2.0-   0.7}$
&  45.7$^{+   2.2+   0.8}_{-   2.1-   0.8}$
&  49.2$^{+   2.3+   0.9}_{-   2.2-   0.9}$
&  48.6$^{+   2.4+   0.8}_{-   2.2-   0.8}$
&  54.3$^{+   2.2+   1.0}_{-   2.2-   1.0}$
&  70.3$^{+   3.7+   1.2}_{-   3.6-   1.2}$
&  74.9$^{+   3.9+   1.3}_{-   3.7-   1.3}$
 \\
    1.17--    1.36
&  16.1$^{+   1.3+   0.3}_{-   1.2-   0.3}$
&  20.8$^{+   1.4+   0.4}_{-   1.3-   0.4}$
&  30.0$^{+   1.6+   0.5}_{-   1.5-   0.5}$
&  30.5$^{+   1.4+   0.5}_{-   1.3-   0.5}$
&  31.1$^{+   1.7+   0.6}_{-   1.7-   0.6}$
&  35.6$^{+   1.8+   0.6}_{-   1.7-   0.6}$
&  35.8$^{+   1.8+   0.6}_{-   1.7-   0.6}$
&  43.0$^{+   2.0+   0.8}_{-   1.9-   0.8}$
&  45.7$^{+   2.1+   0.8}_{-   2.0-   0.8}$
&  48.8$^{+   2.0+   0.9}_{-   1.9-   0.9}$
&  56.6$^{+   3.1+   1.0}_{-   3.0-   1.0}$
&  71.7$^{+   3.5+   1.3}_{-   3.4-   1.3}$
 \\
    1.36--    1.58
&  14.8$^{+   1.1+   0.3}_{-   1.0-   0.3}$
&  18.3$^{+   1.2+   0.3}_{-   1.2-   0.3}$
&  22.8$^{+   1.3+   0.4}_{-   1.2-   0.4}$
&  27.8$^{+   1.2+   0.5}_{-   1.2-   0.5}$
&  24.9$^{+   1.4+   0.4}_{-   1.4-   0.4}$
&  31.0$^{+   1.6+   0.6}_{-   1.5-   0.6}$
&  35.2$^{+   1.7+   0.6}_{-   1.6-   0.6}$
&  36.4$^{+   1.7+   0.7}_{-   1.6-   0.7}$
&  43.1$^{+   1.9+   0.8}_{-   1.8-   0.8}$
&  46.4$^{+   1.8+   0.8}_{-   1.7-   0.8}$
&  56.2$^{+   2.9+   1.0}_{-   2.7-   1.0}$
&  58.5$^{+   3.0+   1.1}_{-   2.8-   1.1}$
 \\
    1.58--    1.85
&  13.8$^{+   1.0+   0.3}_{-   0.9-   0.3}$
&  15.3$^{+   1.0+   0.3}_{-   1.0-   0.3}$
&  21.4$^{+   1.2+   0.4}_{-   1.1-   0.4}$
&  21.8$^{+   1.0+   0.4}_{-   1.0-   0.4}$
&  25.2$^{+   1.3+   0.5}_{-   1.3-   0.5}$
&  26.4$^{+   1.4+   0.5}_{-   1.3-   0.5}$
&  27.8$^{+   1.4+   0.5}_{-   1.3-   0.5}$
&  33.3$^{+   1.5+   0.6}_{-   1.5-   0.6}$
&  36.1$^{+   1.6+   0.7}_{-   1.5-   0.7}$
&  39.8$^{+   1.5+   0.7}_{-   1.5-   0.7}$
&  48.1$^{+   2.5+   0.9}_{-   2.3-   0.9}$
&  50.9$^{+   2.6+   0.9}_{-   2.4-   0.9}$
 \\
    1.85--    2.15
&  12.8$^{+   0.9+   0.2}_{-   0.8-   0.2}$
&  14.2$^{+   0.9+   0.3}_{-   0.9-   0.3}$
&  18.3$^{+   1.0+   0.3}_{-   0.9-   0.3}$
&  18.6$^{+   0.9+   0.3}_{-   0.8-   0.3}$
&  21.4$^{+   1.1+   0.4}_{-   1.1-   0.4}$
&  23.1$^{+   1.2+   0.4}_{-   1.1-   0.4}$
&  25.4$^{+   1.2+   0.5}_{-   1.2-   0.5}$
&  27.6$^{+   1.3+   0.5}_{-   1.2-   0.5}$
&  28.9$^{+   1.3+   0.5}_{-   1.3-   0.5}$
&  30.9$^{+   1.2+   0.6}_{-   1.2-   0.6}$
&  41.7$^{+   2.1+   0.8}_{-   2.0-   0.8}$
&  41.5$^{+   2.1+   0.8}_{-   2.0-   0.8}$
 \\
    2.15--    2.51
&  11.3$^{+   0.8+   0.2}_{-   0.7-   0.2}$
&  12.2$^{+   0.8+   0.2}_{-   0.8-   0.2}$
&  17.2$^{+   0.9+   0.3}_{-   0.9-   0.3}$
&  19.6$^{+   0.8+   0.4}_{-   0.8-   0.4}$
&  20.6$^{+   1.0+   0.4}_{-   1.0-   0.4}$
&  22.3$^{+   1.1+   0.4}_{-   1.0-   0.4}$
&  23.0$^{+   1.1+   0.4}_{-   1.0-   0.4}$
&  25.7$^{+   1.1+   0.5}_{-   1.1-   0.5}$
&  26.8$^{+   1.2+   0.5}_{-   1.1-   0.5}$
&  28.6$^{+   1.1+   0.6}_{-   1.1-   0.6}$
&  33.8$^{+   1.8+   0.7}_{-   1.7-   0.7}$
&  40.3$^{+   2.0+   0.8}_{-   1.9-   0.8}$
 \\
    2.51--    2.93
&  21.8$^{+   1.0+   0.4}_{-   0.9-   0.4}$
&  20.8$^{+   1.0+   0.4}_{-   0.9-   0.4}$
&  31.3$^{+   1.1+   0.6}_{-   1.1-   0.6}$
&  38.8$^{+   1.1+   0.8}_{-   1.0-   0.8}$
&  39.8$^{+   1.3+   0.8}_{-   1.3-   0.8}$
&  39.7$^{+   1.3+   0.8}_{-   1.3-   0.8}$
&  44.7$^{+   1.4+   0.9}_{-   1.3-   0.9}$
&  42.1$^{+   1.3+   0.8}_{-   1.3-   0.8}$
&  36.8$^{+   1.3+   0.7}_{-   1.2-   0.7}$
&  36.1$^{+   1.2+   0.7}_{-   1.1-   0.7}$
&  40.2$^{+   1.8+   0.8}_{-   1.7-   0.8}$
&  45.2$^{+   1.9+   0.9}_{-   1.8-   0.9}$
 \\
    2.93--    3.41
&  65.6$^{+   1.5+   1.3}_{-   1.5-   1.3}$
&  58.4$^{+   1.4+   1.2}_{-   1.4-   1.2}$
&  75.6$^{+   1.6+   1.6}_{-   1.5-   1.6}$
&  86.1$^{+   1.4+   1.8}_{-   1.4-   1.8}$
&  86.2$^{+   1.8+   1.8}_{-   1.7-   1.8}$
&  82.6$^{+   1.7+   1.7}_{-   1.7-   1.7}$
&  87.8$^{+   1.7+   1.8}_{-   1.7-   1.8}$
&  89.6$^{+   1.8+   1.8}_{-   1.7-   1.8}$
&  73.5$^{+   1.6+   1.5}_{-   1.6-   1.5}$
&  71.1$^{+   1.5+   1.5}_{-   1.4-   1.5}$
&  67.6$^{+   2.0+   1.5}_{-   2.0-   1.5}$
&  68.2$^{+   2.1+   1.5}_{-   2.0-   1.5}$
 \\
    3.41--    3.98
&  102.$^{+    2.+    2.}_{-    2.-    2.}$
&  93.0$^{+   1.7+   1.9}_{-   1.6-   1.9}$
&  101.$^{+    2.+    2.}_{-    2.-    2.}$
&  107.$^{+    1.+    2.}_{-    1.-    2.}$
&  107.$^{+    2.+    2.}_{-    2.-    2.}$
&  106.$^{+    2.+    2.}_{-    2.-    2.}$
&  108.$^{+    2.+    2.}_{-    2.-    2.}$
&  107.$^{+    2.+    2.}_{-    2.-    2.}$
&  98.6$^{+   1.8+   2.0}_{-   1.7-   2.0}$
&  96.5$^{+   1.6+   2.0}_{-   1.5-   2.0}$
&  88.4$^{+   2.2+   1.8}_{-   2.1-   1.8}$
&  91.2$^{+   2.3+   1.9}_{-   2.2-   1.9}$
 \\
    3.98--    4.64
&  93.6$^{+   1.6+   2.0}_{-   1.5-   2.0}$
&  88.5$^{+   1.5+   1.9}_{-   1.5-   1.9}$
&  92.5$^{+   1.5+   2.0}_{-   1.5-   2.0}$
&  92.2$^{+   1.3+   1.9}_{-   1.3-   1.9}$
&  88.9$^{+   1.5+   1.9}_{-   1.5-   1.9}$
&  89.4$^{+   1.5+   1.9}_{-   1.5-   1.9}$
&  90.8$^{+   1.5+   1.9}_{-   1.5-   1.9}$
&  88.0$^{+   1.5+   1.9}_{-   1.5-   1.9}$
&  87.9$^{+   1.5+   1.9}_{-   1.5-   1.9}$
&  86.9$^{+   1.4+   1.8}_{-   1.4-   1.8}$
&  80.3$^{+   1.9+   1.7}_{-   1.9-   1.7}$
&  75.4$^{+   1.9+   1.6}_{-   1.9-   1.6}$
 \\
    4.64--    5.41
&  74.7$^{+   1.3+   1.6}_{-   1.3-   1.6}$
&  72.7$^{+   1.3+   1.6}_{-   1.3-   1.6}$
&  71.4$^{+   1.2+   1.6}_{-   1.2-   1.6}$
&  69.3$^{+   1.0+   1.5}_{-   1.0-   1.5}$
&  68.5$^{+   1.3+   1.5}_{-   1.2-   1.5}$
&  68.4$^{+   1.2+   1.5}_{-   1.2-   1.5}$
&  69.3$^{+   1.2+   1.5}_{-   1.2-   1.5}$
&  68.3$^{+   1.2+   1.5}_{-   1.2-   1.5}$
&  69.4$^{+   1.3+   1.5}_{-   1.2-   1.5}$
&  69.6$^{+   1.2+   1.5}_{-   1.1-   1.5}$
&  63.1$^{+   1.6+   1.4}_{-   1.6-   1.4}$
&  63.3$^{+   1.6+   1.4}_{-   1.6-   1.4}$
 \\
    5.41--    6.31
&  54.8$^{+   1.0+   1.3}_{-   1.0-   1.3}$
&  55.8$^{+   1.0+   1.3}_{-   1.0-   1.3}$
&  53.7$^{+   1.0+   1.3}_{-   1.0-   1.3}$
&  52.0$^{+   0.8+   1.2}_{-   0.8-   1.2}$
&  53.3$^{+   1.0+   1.2}_{-   1.0-   1.2}$
&  53.7$^{+   1.0+   1.2}_{-   1.0-   1.2}$
&  51.3$^{+   1.0+   1.2}_{-   1.0-   1.2}$
&  51.7$^{+   1.0+   1.2}_{-   1.0-   1.2}$
&  50.5$^{+   1.0+   1.2}_{-   1.0-   1.2}$
&  50.4$^{+   0.9+   1.2}_{-   0.9-   1.2}$
&  49.6$^{+   1.3+   1.1}_{-   1.3-   1.1}$
&  47.9$^{+   1.3+   1.1}_{-   1.3-   1.1}$
 \\
    6.31--    7.36
&  40.6$^{+   0.8+   1.0}_{-   0.8-   1.0}$
&  41.0$^{+   0.8+   1.0}_{-   0.8-   1.0}$
&  39.9$^{+   0.8+   1.0}_{-   0.8-   1.0}$
&  39.7$^{+   0.7+   0.9}_{-   0.7-   0.9}$
&  39.7$^{+   0.8+   0.9}_{-   0.8-   0.9}$
&  40.0$^{+   0.8+   1.0}_{-   0.8-   1.0}$
&  38.1$^{+   0.8+   0.9}_{-   0.8-   0.9}$
&  39.3$^{+   0.8+   0.9}_{-   0.8-   0.9}$
&  37.5$^{+   0.8+   0.9}_{-   0.8-   0.9}$
&  37.6$^{+   0.7+   0.9}_{-   0.7-   0.9}$
&  35.8$^{+   1.0+   0.9}_{-   1.0-   0.9}$
&  34.7$^{+   1.0+   0.8}_{-   1.0-   0.8}$
 \\
    7.36--    8.58
&  29.4$^{+   0.7+   0.9}_{-   0.6-   0.9}$
&  29.8$^{+   0.7+   0.9}_{-   0.6-   0.9}$
&  28.1$^{+   0.6+   0.8}_{-   0.6-   0.8}$
&  29.2$^{+   0.5+   0.9}_{-   0.5-   0.9}$
&  28.6$^{+   0.7+   0.8}_{-   0.6-   0.8}$
&  28.0$^{+   0.6+   0.8}_{-   0.6-   0.8}$
&  27.6$^{+   0.6+   0.8}_{-   0.6-   0.8}$
&  27.4$^{+   0.6+   0.8}_{-   0.6-   0.8}$
&  27.4$^{+   0.6+   0.8}_{-   0.6-   0.8}$
&  28.4$^{+   0.6+   0.8}_{-   0.6-   0.8}$
&  25.9$^{+   0.8+   0.6}_{-   0.8-   0.6}$
&  26.8$^{+   0.9+   0.7}_{-   0.8-   0.7}$
 \\
    8.58--   10.00
&  19.9$^{+   0.3+   0.6}_{-   0.3-   0.6}$
&  19.9$^{+   0.3+   0.6}_{-   0.3-   0.6}$
&  18.9$^{+   0.2+   0.6}_{-   0.2-   0.6}$
&  19.7$^{+   0.2+   0.6}_{-   0.2-   0.6}$
&  19.0$^{+   0.3+   0.6}_{-   0.3-   0.6}$
&  19.3$^{+   0.3+   0.6}_{-   0.3-   0.6}$
&  19.0$^{+   0.3+   0.6}_{-   0.3-   0.6}$
&  18.7$^{+   0.3+   0.6}_{-   0.3-   0.6}$
&  18.5$^{+   0.3+   0.6}_{-   0.3-   0.6}$
&  18.5$^{+   0.2+   0.6}_{-   0.2-   0.6}$
&  17.1$^{+   0.3+   0.5}_{-   0.3-   0.5}$
&  17.6$^{+   0.3+   0.5}_{-   0.3-   0.5}$
 \\
      \hline
      \hline
    \end{tabular}
}
  \end{center}
}
\end{table}

\begin{table}[b!]
\tiny{
  \caption
{Observed helium fluxes}
  \label{tab:sumhe1}
\renewcommand{\arraystretch}{1.5}
  \begin{center}
\rotatebox{90}{
    \begin{tabular}{cllllllllllll}
      \hline
      \hline
      \begin{tabular}{@{}c@{}}       
      Energy range\\
      (GeV/n)
      \end{tabular}
      & 
      \multicolumn{12}{c}{
      \begin{tabular}{@{}c@{}}       
	Flux$\pm~\Delta$Flux$_{\rm sta}
	\pm~\Delta$Flux$_{\rm sys}$\\
	(m$^{~-2}$sr$^{~-1}$s$^{~-1}$(GeV/n)$^{~-1}$)
      \end{tabular}
      }
      \\
      \hline
      &  \multicolumn{12}{c}{Atmospheric depth range [mean] (g/cm$^2$)}\\
      &  \multicolumn{1}{c}{4.46--4.80}  
      &  \multicolumn{1}{c}{4.86--7.21}  
      &  \multicolumn{1}{c}{7.04--8.23}  
      &  \multicolumn{1}{c}{8.24--9.08}  
      &  \multicolumn{1}{c}{9.06--9.54}  
      &  \multicolumn{1}{c}{9.60--11.4} 
      &  \multicolumn{1}{c}{11.4--12.5}
      &  \multicolumn{1}{c}{12.7--14.6}  
      &  \multicolumn{1}{c}{14.7--16.4} 
      &  \multicolumn{1}{c}{16.5--19.8}   
      &  \multicolumn{1}{c}{21.2--25.0}   
      &  \multicolumn{1}{c}{25.0--28.2}   \\
       & \multicolumn{1}{c}{[4.58]} & \multicolumn{1}{c}{[5.82]} 
      & \multicolumn{1}{c}{[7.81]} & \multicolumn{1}{c}{[8.70]} 
      & \multicolumn{1}{c}{[9.33]} & \multicolumn{1}{c}{[10.4]} 
      & \multicolumn{1}{c}{[11.9]} & \multicolumn{1}{c}{[13.6]} 
      & \multicolumn{1}{c}{[15.6]}& \multicolumn{1}{c}{[17.6]}
      & \multicolumn{1}{c}{[23.4]}& \multicolumn{1}{c}{[26.4]}\\
      \hline 
    0.46--    0.54
&$<$ 0.310
&.237$^{+.414+.007}_{-.149-.007}$
&.215$^{+.376+.006}_{-.135-.006}$
&.160$^{+.279+.004}_{-.100-.004}$
&.246$^{+.430+.007}_{-.155-.007}$
&.476$^{+.535+.013}_{-.300-.013}$
&.473$^{+.532+.013}_{-.298-.013}$
&$<$ 0.309
&$<$ 0.319
&$<$ 0.263
&$<$ 0.558
&.443$^{+.775+.012}_{-.279-.012}$
 \\
    0.54--    0.63
&$<$ 0.268
&$<$ 0.264
&.371$^{+.418+.010}_{-.234-.010}$
&.551$^{+.382+.015}_{-.229-.015}$
&.425$^{+.478+.012}_{-.267-.012}$
&.206$^{+.360+.006}_{-.130-.006}$
&.818$^{+.567+.023}_{-.340-.023}$
&.414$^{+.466+.011}_{-.261-.011}$
&.214$^{+.374+.006}_{-.135-.006}$
&.353$^{+.397+.010}_{-.222-.010}$
&.374$^{+.654+.010}_{-.235-.010}$
&.383$^{+.670+.011}_{-.241-.011}$
 \\
    0.63--    0.74
&.539$^{+.413+.015}_{-.340-.015}$
&.354$^{+.398+.010}_{-.223-.010}$
&.321$^{+.361+.009}_{-.202-.009}$
&  1.07$^{+  0.45+  0.03}_{-  0.32-  0.03}$
&  1.83$^{+  0.70+  0.05}_{-  0.59-  0.05}$
&  1.07$^{+  0.58+  0.03}_{-  0.39-  0.03}$
&  1.94$^{+  0.67+  0.05}_{-  0.56-  0.05}$
&  2.51$^{+  0.77+  0.07}_{-  0.66-  0.07}$
&  1.29$^{+  0.61+  0.04}_{-  0.51-  0.04}$
&.915$^{+.498+.025}_{-.331-.025}$
&.646$^{+.727+.018}_{-.407-.018}$
&  2.65$^{+  1.09+  0.07}_{-  0.89-  0.07}$
 \\
    0.74--    0.86
&.466$^{+.357+.013}_{-.293-.013}$
&  1.68$^{+  0.58+  0.05}_{-  0.49-  0.05}$
&  1.66$^{+  0.59+  0.05}_{-  0.44-  0.05}$
&  1.65$^{+  0.49+  0.05}_{-  0.38-  0.05}$
&  2.85$^{+  0.76+  0.08}_{-  0.66-  0.08}$
&  1.38$^{+  0.58+  0.04}_{-  0.41-  0.04}$
&  1.99$^{+  0.66+  0.06}_{-  0.57-  0.06}$
&  2.32$^{+  0.67+  0.06}_{-  0.57-  0.06}$
&  2.07$^{+  0.69+  0.06}_{-  0.59-  0.06}$
&  2.37$^{+  0.63+  0.07}_{-  0.55-  0.07}$
&  1.95$^{+  0.92+  0.05}_{-  0.76-  0.05}$
&  2.00$^{+  0.94+  0.06}_{-  0.78-  0.06}$
 \\
    0.86--    1.00
&  1.88$^{+  0.58+  0.05}_{-  0.49-  0.05}$
&  1.98$^{+  0.57+  0.06}_{-  0.49-  0.06}$
&  3.47$^{+  0.70+  0.10}_{-  0.62-  0.10}$
&  4.45$^{+  0.69+  0.13}_{-  0.64-  0.13}$
&  5.07$^{+  0.93+  0.14}_{-  0.85-  0.14}$
&  4.64$^{+  0.84+  0.13}_{-  0.75-  0.13}$
&  6.73$^{+  1.03+  0.19}_{-  0.95-  0.19}$
&  4.54$^{+  0.84+  0.13}_{-  0.76-  0.13}$
&  3.86$^{+  0.80+  0.11}_{-  0.71-  0.11}$
&  4.10$^{+  0.77+  0.12}_{-  0.64-  0.12}$
&  2.41$^{+  0.92+  0.07}_{-  0.78-  0.07}$
&  4.20$^{+  1.19+  0.12}_{-  1.04-  0.12}$
 \\
    1.00--    1.17
&  6.60$^{+  0.96+  0.19}_{-  0.89-  0.19}$
&  7.30$^{+  1.00+  0.21}_{-  0.87-  0.21}$
&  10.0$^{+   1.1+   0.3}_{-   1.0-   0.3}$
&  10.9$^{+   0.9+   0.3}_{-   0.9-   0.3}$
&  12.9$^{+   1.3+   0.4}_{-   1.2-   0.4}$
&  12.6$^{+   1.3+   0.4}_{-   1.2-   0.4}$
&  13.2$^{+   1.3+   0.4}_{-   1.2-   0.4}$
&  12.7$^{+   1.3+   0.4}_{-   1.2-   0.4}$
&  9.16$^{+  1.11+  0.26}_{-  1.03-  0.26}$
&  7.47$^{+  0.92+  0.21}_{-  0.85-  0.21}$
&  10.2$^{+   1.6+   0.3}_{-   1.4-   0.3}$
&  9.81$^{+  1.56+  0.28}_{-  1.42-  0.28}$
 \\
    1.17--    1.36
&  19.8$^{+   1.5+   0.6}_{-   1.4-   0.6}$
&  20.4$^{+   1.5+   0.6}_{-   1.4-   0.6}$
&  21.5$^{+   1.5+   0.6}_{-   1.4-   0.6}$
&  22.8$^{+   1.3+   0.7}_{-   1.2-   0.7}$
&  26.1$^{+   1.7+   0.7}_{-   1.6-   0.7}$
&  23.3$^{+   1.6+   0.7}_{-   1.5-   0.7}$
&  23.8$^{+   1.6+   0.7}_{-   1.5-   0.7}$
&  21.1$^{+   1.6+   0.6}_{-   1.4-   0.6}$
&  18.3$^{+   1.5+   0.5}_{-   1.4-   0.5}$
&  17.7$^{+   1.3+   0.5}_{-   1.2-   0.5}$
&  15.6$^{+   1.8+   0.4}_{-   1.6-   0.4}$
&  14.7$^{+   1.7+   0.4}_{-   1.6-   0.4}$
 \\
    1.36--    1.58
&  27.3$^{+   1.6+   0.8}_{-   1.5-   0.8}$
&  27.7$^{+   1.6+   0.8}_{-   1.5-   0.8}$
&  29.3$^{+   1.6+   0.8}_{-   1.5-   0.8}$
&  28.0$^{+   1.3+   0.8}_{-   1.3-   0.8}$
&  31.3$^{+   1.7+   0.9}_{-   1.7-   0.9}$
&  29.1$^{+   1.7+   0.8}_{-   1.6-   0.8}$
&  27.2$^{+   1.6+   0.8}_{-   1.5-   0.8}$
&  24.0$^{+   1.5+   0.7}_{-   1.4-   0.7}$
&  27.7$^{+   1.7+   0.8}_{-   1.6-   0.8}$
&  22.8$^{+   1.4+   0.7}_{-   1.3-   0.7}$
&  23.9$^{+   2.1+   0.7}_{-   1.9-   0.7}$
&  18.7$^{+   1.8+   0.5}_{-   1.7-   0.5}$
 \\
    1.58--    1.85
&  27.4$^{+   1.5+   0.8}_{-   1.4-   0.8}$
&  28.5$^{+   1.5+   0.8}_{-   1.4-   0.8}$
&  25.9$^{+   1.4+   0.7}_{-   1.3-   0.7}$
&  27.9$^{+   1.2+   0.8}_{-   1.2-   0.8}$
&  26.6$^{+   1.5+   0.8}_{-   1.4-   0.8}$
&  25.7$^{+   1.4+   0.7}_{-   1.4-   0.7}$
&  22.8$^{+   1.4+   0.7}_{-   1.3-   0.7}$
&  23.3$^{+   1.4+   0.7}_{-   1.3-   0.7}$
&  25.6$^{+   1.5+   0.7}_{-   1.4-   0.7}$
&  23.6$^{+   1.3+   0.7}_{-   1.2-   0.7}$
&  22.1$^{+   1.9+   0.6}_{-   1.7-   0.6}$
&  20.4$^{+   1.8+   0.6}_{-   1.7-   0.6}$
 \\
    1.85--    2.15
&  23.9$^{+   1.3+   0.7}_{-   1.2-   0.7}$
&  23.6$^{+   1.3+   0.7}_{-   1.2-   0.7}$
&  22.3$^{+   1.2+   0.6}_{-   1.1-   0.6}$
&  23.8$^{+   1.0+   0.7}_{-   1.0-   0.7}$
&  23.6$^{+   1.3+   0.7}_{-   1.2-   0.7}$
&  21.9$^{+   1.2+   0.6}_{-   1.2-   0.6}$
&  19.6$^{+   1.2+   0.6}_{-   1.1-   0.6}$
&  19.7$^{+   1.2+   0.6}_{-   1.1-   0.6}$
&  19.3$^{+   1.2+   0.6}_{-   1.1-   0.6}$
&  18.3$^{+   1.1+   0.5}_{-   1.0-   0.5}$
&  15.1$^{+   1.4+   0.4}_{-   1.3-   0.4}$
&  15.8$^{+   1.5+   0.5}_{-   1.4-   0.5}$
 \\
    2.15--    2.51
&  19.2$^{+   1.1+   0.6}_{-   1.0-   0.6}$
&  19.3$^{+   1.1+   0.6}_{-   1.0-   0.6}$
&  18.0$^{+   1.0+   0.5}_{-   0.9-   0.5}$
&  18.3$^{+   0.9+   0.5}_{-   0.8-   0.5}$
&  17.4$^{+   1.0+   0.5}_{-   1.0-   0.5}$
&  16.4$^{+   1.0+   0.5}_{-   0.9-   0.5}$
&  17.1$^{+   1.0+   0.5}_{-   1.0-   0.5}$
&  13.8$^{+   0.9+   0.4}_{-   0.9-   0.4}$
&  14.6$^{+   1.0+   0.4}_{-   0.9-   0.4}$
&  15.5$^{+   0.9+   0.5}_{-   0.9-   0.5}$
&  13.8$^{+   1.2+   0.4}_{-   1.2-   0.4}$
&  11.5$^{+   1.2+   0.3}_{-   1.1-   0.3}$
 \\
    2.51--    2.93
&  16.1$^{+   0.9+   0.5}_{-   0.9-   0.5}$
&  16.1$^{+   0.9+   0.5}_{-   0.9-   0.5}$
&  13.2$^{+   0.8+   0.4}_{-   0.7-   0.4}$
&  14.6$^{+   0.7+   0.4}_{-   0.7-   0.4}$
&  14.0$^{+   0.9+   0.4}_{-   0.8-   0.4}$
&  15.4$^{+   0.9+   0.5}_{-   0.9-   0.5}$
&  12.9$^{+   0.8+   0.4}_{-   0.8-   0.4}$
&  12.4$^{+   0.8+   0.4}_{-   0.8-   0.4}$
&  12.7$^{+   0.8+   0.4}_{-   0.8-   0.4}$
&  11.4$^{+   0.7+   0.3}_{-   0.7-   0.3}$
&  11.4$^{+   1.1+   0.3}_{-   1.0-   0.3}$
&  9.64$^{+  0.99+  0.28}_{-  0.89-  0.28}$
 \\
    2.93--    3.41
&  11.8$^{+   0.7+   0.4}_{-   0.7-   0.4}$
&  11.3$^{+   0.7+   0.3}_{-   0.7-   0.3}$
&  10.9$^{+   0.7+   0.3}_{-   0.6-   0.3}$
&  10.4$^{+   0.6+   0.3}_{-   0.5-   0.3}$
&  9.84$^{+  0.68+  0.29}_{-  0.64-  0.29}$
&  10.1$^{+   0.7+   0.3}_{-   0.6-   0.3}$
&  9.89$^{+  0.67+  0.29}_{-  0.63-  0.29}$
&  10.1$^{+   0.7+   0.3}_{-   0.6-   0.3}$
&  9.86$^{+  0.69+  0.29}_{-  0.64-  0.29}$
&  8.98$^{+  0.59+  0.27}_{-  0.56-  0.27}$
&  7.37$^{+  0.80+  0.22}_{-  0.71-  0.22}$
&  7.85$^{+  0.82+  0.23}_{-  0.77-  0.23}$
 \\
    3.41--    3.98
&  8.90$^{+  0.60+  0.27}_{-  0.56-  0.27}$
&  8.10$^{+  0.57+  0.24}_{-  0.53-  0.24}$
&  8.45$^{+  0.55+  0.25}_{-  0.52-  0.25}$
&  8.85$^{+  0.48+  0.26}_{-  0.46-  0.26}$
&  8.18$^{+  0.58+  0.24}_{-  0.54-  0.24}$
&  8.03$^{+  0.57+  0.24}_{-  0.53-  0.24}$
&  8.23$^{+  0.57+  0.25}_{-  0.53-  0.25}$
&  7.07$^{+  0.53+  0.21}_{-  0.50-  0.21}$
&  6.28$^{+  0.51+  0.19}_{-  0.48-  0.19}$
&  6.50$^{+  0.47+  0.19}_{-  0.44-  0.19}$
&  5.52$^{+  0.62+  0.17}_{-  0.58-  0.17}$
&  5.33$^{+  0.64+  0.16}_{-  0.60-  0.16}$
 \\
    3.98--    4.64
&  7.45$^{+  0.51+  0.22}_{-  0.48-  0.22}$
&  6.04$^{+  0.46+  0.18}_{-  0.42-  0.18}$
&  5.97$^{+  0.43+  0.18}_{-  0.40-  0.18}$
&  6.17$^{+  0.37+  0.19}_{-  0.35-  0.19}$
&  6.67$^{+  0.49+  0.20}_{-  0.45-  0.20}$
&  6.25$^{+  0.46+  0.19}_{-  0.43-  0.19}$
&  6.76$^{+  0.48+  0.20}_{-  0.45-  0.20}$
&  5.42$^{+  0.44+  0.16}_{-  0.40-  0.16}$
&  5.18$^{+  0.43+  0.16}_{-  0.40-  0.16}$
&  5.70$^{+  0.41+  0.17}_{-  0.38-  0.17}$
&  4.43$^{+  0.54+  0.13}_{-  0.47-  0.13}$
&  4.31$^{+  0.52+  0.13}_{-  0.49-  0.13}$
 \\
    4.64--    5.41
&  4.36$^{+  0.18+  0.14}_{-  0.18-  0.14}$
&  4.68$^{+  0.19+  0.15}_{-  0.18-  0.15}$
&  4.14$^{+  0.12+  0.13}_{-  0.12-  0.13}$
&  4.38$^{+  0.15+  0.14}_{-  0.14-  0.14}$
&  4.21$^{+  0.18+  0.13}_{-  0.17-  0.13}$
&  4.46$^{+  0.18+  0.14}_{-  0.18-  0.14}$
&  4.14$^{+  0.18+  0.13}_{-  0.17-  0.13}$
&  3.74$^{+  0.17+  0.12}_{-  0.16-  0.12}$
&  3.57$^{+  0.17+  0.11}_{-  0.16-  0.11}$
&  3.74$^{+  0.16+  0.12}_{-  0.15-  0.12}$
&  3.06$^{+  0.21+  0.10}_{-  0.20-  0.10}$
&  2.66$^{+  0.20+  0.09}_{-  0.19-  0.09}$
 \\
    5.41--    6.31
&  3.03$^{+  0.14+  0.09}_{-  0.13-  0.09}$
&  3.18$^{+  0.14+  0.10}_{-  0.14-  0.10}$
&  3.01$^{+  0.10+  0.09}_{-  0.09-  0.09}$
&  2.99$^{+  0.11+  0.09}_{-  0.11-  0.09}$
&  3.22$^{+  0.15+  0.10}_{-  0.14-  0.10}$
&  3.11$^{+  0.14+  0.10}_{-  0.13-  0.10}$
&  2.90$^{+  0.14+  0.09}_{-  0.13-  0.09}$
&  2.81$^{+  0.13+  0.09}_{-  0.13-  0.09}$
&  2.67$^{+  0.13+  0.08}_{-  0.13-  0.08}$
&  2.50$^{+  0.12+  0.08}_{-  0.11-  0.08}$
&  2.13$^{+  0.16+  0.07}_{-  0.15-  0.07}$
&  2.15$^{+  0.16+  0.07}_{-  0.15-  0.07}$
 \\
    6.31--    7.36
&  2.38$^{+  0.11+  0.07}_{-  0.11-  0.07}$
&  2.22$^{+  0.11+  0.07}_{-  0.11-  0.07}$
&  2.23$^{+  0.08+  0.07}_{-  0.07-  0.07}$
&  2.22$^{+  0.09+  0.07}_{-  0.09-  0.07}$
&  2.16$^{+  0.11+  0.07}_{-  0.11-  0.07}$
&  2.16$^{+  0.11+  0.07}_{-  0.10-  0.07}$
&  1.90$^{+  0.10+  0.06}_{-  0.10-  0.06}$
&  1.83$^{+  0.10+  0.06}_{-  0.10-  0.06}$
&  1.84$^{+  0.10+  0.06}_{-  0.10-  0.06}$
&  1.85$^{+  0.09+  0.06}_{-  0.09-  0.06}$
&  1.60$^{+  0.13+  0.05}_{-  0.12-  0.05}$
&  1.55$^{+  0.13+  0.05}_{-  0.12-  0.05}$
 \\
    7.36--    8.58
&  1.73$^{+  0.09+  0.05}_{-  0.09-  0.05}$
&  1.60$^{+  0.09+  0.05}_{-  0.08-  0.05}$
&  1.51$^{+  0.06+  0.05}_{-  0.06-  0.05}$
&  1.54$^{+  0.07+  0.05}_{-  0.07-  0.05}$
&  1.66$^{+  0.09+  0.05}_{-  0.09-  0.05}$
&  1.54$^{+  0.09+  0.05}_{-  0.08-  0.05}$
&  1.49$^{+  0.08+  0.05}_{-  0.08-  0.05}$
&  1.31$^{+  0.08+  0.04}_{-  0.08-  0.04}$
&  1.33$^{+  0.08+  0.04}_{-  0.08-  0.04}$
&  1.42$^{+  0.08+  0.04}_{-  0.07-  0.04}$
&  1.17$^{+  0.10+  0.04}_{-  0.10-  0.04}$
&  1.00$^{+  0.09+  0.03}_{-  0.09-  0.03}$
 \\
    8.58--   10.00
&  1.04$^{+  0.07+  0.03}_{-  0.06-  0.03}$
&  1.07$^{+  0.07+  0.03}_{-  0.06-  0.03}$
&  1.07$^{+  0.05+  0.03}_{-  0.04-  0.03}$
&  1.08$^{+  0.05+  0.03}_{-  0.05-  0.03}$
&  1.07$^{+  0.07+  0.03}_{-  0.06-  0.03}$
&  1.06$^{+  0.07+  0.03}_{-  0.06-  0.03}$
&.983$^{+.064+.031}_{-.060-.031}$
&.999$^{+.065+.031}_{-.061-.031}$
&.886$^{+.062+.028}_{-.058-.028}$
&.961$^{+.058+.030}_{-.055-.030}$
&.808$^{+.079+.025}_{-.071-.025}$
&.814$^{+.077+.026}_{-.073-.026}$
 \\
      \hline
      \hline
    \end{tabular}
}
  \end{center}
}
\end{table}

\begin{table}[b!]
\tiny{
  \caption
{Observed negative muon fluxes}
  \label{tab:summm1}
\renewcommand{\arraystretch}{1.5}
  \begin{center}
\rotatebox{90}{
    \begin{tabular}{cllllllllllll}
      \hline
      \hline
      \begin{tabular}{@{}c@{}}       
      Momentum range\\
      (GeV/$c$)
      \end{tabular}
      & 
      \multicolumn{12}{c}{
      \begin{tabular}{@{}c@{}}       
	Flux$\pm~\Delta$Flux$_{\rm sta}
	\pm~\Delta$Flux$_{\rm sys}$\\
	(m$^{~-2}$sr$^{~-1}$s$^{~-1}$(GeV/$c$)$^{~-1}$)
      \end{tabular}
      }
      \\
      \hline
      &       \multicolumn{12}{c}{Atmospheric depth range [mean] (g/cm$^2$)}\\
      &  \multicolumn{1}{c}{4.46--4.80}  
      &  \multicolumn{1}{c}{4.86--7.21}  
      &  \multicolumn{1}{c}{7.04--8.23}  
      &  \multicolumn{1}{c}{8.24--9.08}  
      &  \multicolumn{1}{c}{9.06--9.54}  
      &  \multicolumn{1}{c}{9.60--11.4} 
      &  \multicolumn{1}{c}{11.4--12.5}
      &  \multicolumn{1}{c}{12.7--14.6}  
      &  \multicolumn{1}{c}{14.7--16.4} 
      &  \multicolumn{1}{c}{16.5--19.8}   
      &  \multicolumn{1}{c}{21.2--25.0}   
      &  \multicolumn{1}{c}{25.0--28.2}   \\
      & \multicolumn{1}{c}{[4.58]} & \multicolumn{1}{c}{[5.82]} 
      & \multicolumn{1}{c}{[7.81]} & \multicolumn{1}{c}{[8.70]} 
      & \multicolumn{1}{c}{[9.33]} & \multicolumn{1}{c}{[10.4]} 
      & \multicolumn{1}{c}{[11.9]} & \multicolumn{1}{c}{[13.6]} 
      & \multicolumn{1}{c}{[15.6]}& \multicolumn{1}{c}{[17.6]}
      & \multicolumn{1}{c}{[23.4]}& \multicolumn{1}{c}{[26.4]}\\
      \hline
    0.50--    0.58
&  6.76$^{+  2.80+  0.32}_{-  2.27-  0.32}$
&  13.4$^{+   4.0+   0.5}_{-   3.1-   0.5}$
&  15.0$^{+   4.0+   0.6}_{-   3.1-   0.6}$
&  22.1$^{+   3.8+   0.9}_{-   3.4-   0.9}$
&  23.2$^{+   5.0+   0.8}_{-   4.5-   0.8}$
&  19.7$^{+   4.6+   0.9}_{-   3.7-   0.9}$
&  15.0$^{+   4.0+   0.5}_{-   3.5-   0.5}$
&  22.1$^{+   4.8+   0.9}_{-   4.3-   0.9}$
&  33.3$^{+   5.7+   1.3}_{-   5.1-   1.3}$
&  26.6$^{+   4.5+   1.1}_{-   4.1-   1.1}$
&  30.8$^{+   7.8+   1.0}_{-   6.9-   1.0}$
&  57.0$^{+  10.2+   1.8}_{-   9.3-   1.8}$
 \\
    0.58--    0.67
&  13.3$^{+   3.5+   0.5}_{-   2.8-   0.5}$
&  14.0$^{+   3.5+   0.4}_{-   3.1-   0.4}$
&  18.0$^{+   3.8+   0.5}_{-   3.1-   0.5}$
&  14.2$^{+   2.8+   0.4}_{-   2.5-   0.4}$
&  22.7$^{+   4.4+   0.6}_{-   3.9-   0.6}$
&  21.3$^{+   4.1+   0.7}_{-   3.7-   0.7}$
&  21.7$^{+   4.2+   0.6}_{-   3.7-   0.6}$
&  26.3$^{+   4.5+   0.8}_{-   4.1-   0.8}$
&  22.8$^{+   4.2+   0.7}_{-   3.8-   0.7}$
&  25.0$^{+   4.0+   0.8}_{-   3.6-   0.8}$
&  33.0$^{+   6.9+   0.8}_{-   6.1-   0.8}$
&  50.7$^{+   8.8+   1.2}_{-   7.4-   1.2}$
 \\
    0.67--    0.78
&  10.7$^{+   2.8+   0.3}_{-   2.2-   0.3}$
&  9.71$^{+  2.59+  0.22}_{-  2.26-  0.22}$
&  14.5$^{+   3.1+   0.3}_{-   2.5-   0.3}$
&  11.9$^{+   2.3+   0.3}_{-   2.0-   0.3}$
&  14.5$^{+   3.2+   0.3}_{-   2.9-   0.3}$
&  15.4$^{+   3.1+   0.4}_{-   2.7-   0.4}$
&  12.7$^{+   2.8+   0.3}_{-   2.5-   0.3}$
&  22.9$^{+   3.9+   0.5}_{-   3.6-   0.5}$
&  23.5$^{+   4.0+   0.6}_{-   3.7-   0.6}$
&  20.1$^{+   3.3+   0.5}_{-   3.0-   0.5}$
&  39.4$^{+   6.6+   0.8}_{-   5.9-   0.8}$
&  31.5$^{+   6.2+   0.7}_{-   5.6-   0.7}$
 \\
    0.78--    0.90
&  5.25$^{+  1.88+  0.13}_{-  1.39-  0.13}$
&  8.38$^{+  2.12+  0.16}_{-  1.84-  0.16}$
&  10.3$^{+   2.3+   0.2}_{-   2.1-   0.2}$
&  11.8$^{+   2.0+   0.2}_{-   1.8-   0.2}$
&  14.2$^{+   2.9+   0.3}_{-   2.6-   0.3}$
&  12.7$^{+   2.5+   0.2}_{-   2.3-   0.2}$
&  13.5$^{+   2.7+   0.3}_{-   2.5-   0.3}$
&  21.1$^{+   3.2+   0.4}_{-   2.9-   0.4}$
&  17.1$^{+   3.1+   0.4}_{-   2.8-   0.4}$
&  21.5$^{+   3.1+   0.4}_{-   2.9-   0.4}$
&  26.1$^{+   5.0+   0.5}_{-   4.5-   0.5}$
&  30.8$^{+   5.5+   0.6}_{-   5.0-   0.6}$
 \\
    0.90--    1.05
&  4.38$^{+  0.73+  0.14}_{-  0.66-  0.14}$
&  6.96$^{+  0.93+  0.20}_{-  0.86-  0.20}$
&  7.26$^{+  0.65+  0.21}_{-  0.62-  0.21}$
&  9.83$^{+  0.88+  0.29}_{-  0.83-  0.29}$
&  9.30$^{+  1.07+  0.27}_{-  1.00-  0.27}$
&  9.32$^{+  1.04+  0.27}_{-  0.97-  0.27}$
&  10.6$^{+   1.1+   0.3}_{-   1.1-   0.3}$
&  13.1$^{+   1.3+   0.4}_{-   1.2-   0.4}$
&  13.6$^{+   1.3+   0.4}_{-   1.2-   0.4}$
&  17.6$^{+   1.4+   0.5}_{-   1.3-   0.5}$
&  21.1$^{+   2.2+   0.6}_{-   2.0-   0.6}$
&  22.6$^{+   2.2+   0.7}_{-   2.1-   0.7}$
 \\
    1.05--    1.21
&  3.69$^{+  0.63+  0.09}_{-  0.57-  0.09}$
&  4.32$^{+  0.66+  0.09}_{-  0.61-  0.09}$
&  6.97$^{+  0.59+  0.15}_{-  0.54-  0.15}$
&  6.23$^{+  0.62+  0.14}_{-  0.58-  0.14}$
&  9.31$^{+  0.95+  0.20}_{-  0.89-  0.20}$
&  9.87$^{+  0.96+  0.22}_{-  0.91-  0.22}$
&  11.2$^{+   1.0+   0.2}_{-   0.9-   0.2}$
&  11.6$^{+   1.1+   0.3}_{-   1.0-   0.3}$
&  12.3$^{+   1.1+   0.3}_{-   1.0-   0.3}$
&  16.1$^{+   1.2+   0.4}_{-   1.1-   0.4}$
&  17.5$^{+   1.7+   0.4}_{-   1.6-   0.4}$
&  19.3$^{+   1.9+   0.4}_{-   1.8-   0.4}$
 \\
    1.21--    1.41
&  3.04$^{+  0.52+  0.06}_{-  0.47-  0.06}$
&  4.52$^{+  0.61+  0.08}_{-  0.57-  0.08}$
&  4.61$^{+  0.43+  0.09}_{-  0.39-  0.09}$
&  6.94$^{+  0.60+  0.13}_{-  0.57-  0.13}$
&  7.08$^{+  0.75+  0.13}_{-  0.70-  0.13}$
&  6.36$^{+  0.72+  0.12}_{-  0.68-  0.12}$
&  8.13$^{+  0.79+  0.15}_{-  0.74-  0.15}$
&  10.0$^{+   0.9+   0.2}_{-   0.8-   0.2}$
&  8.63$^{+  0.82+  0.16}_{-  0.77-  0.16}$
&  11.0$^{+   0.9+   0.2}_{-   0.8-   0.2}$
&  12.5$^{+   1.3+   0.2}_{-   1.2-   0.2}$
&  17.4$^{+   1.6+   0.3}_{-   1.5-   0.3}$
 \\
    1.41--    1.63
&  2.41$^{+  0.40+  0.04}_{-  0.36-  0.04}$
&  3.42$^{+  0.48+  0.06}_{-  0.45-  0.06}$
&  4.42$^{+  0.39+  0.08}_{-  0.36-  0.08}$
&  5.28$^{+  0.48+  0.09}_{-  0.46-  0.09}$
&  4.22$^{+  0.53+  0.07}_{-  0.49-  0.07}$
&  5.42$^{+  0.60+  0.09}_{-  0.56-  0.09}$
&  6.03$^{+  0.63+  0.10}_{-  0.59-  0.10}$
&  7.35$^{+  0.69+  0.13}_{-  0.66-  0.13}$
&  9.81$^{+  0.83+  0.17}_{-  0.77-  0.17}$
&  10.0$^{+   0.8+   0.2}_{-   0.7-   0.2}$
&  10.9$^{+   1.1+   0.2}_{-   1.1-   0.2}$
&  13.5$^{+   1.3+   0.2}_{-   1.2-   0.2}$
 \\
    1.63--    1.90
&  2.68$^{+  0.39+  0.05}_{-  0.36-  0.05}$
&  2.64$^{+  0.38+  0.04}_{-  0.35-  0.04}$
&  3.21$^{+  0.29+  0.05}_{-  0.27-  0.05}$
&  3.79$^{+  0.37+  0.06}_{-  0.35-  0.06}$
&  4.01$^{+  0.47+  0.07}_{-  0.44-  0.07}$
&  4.42$^{+  0.51+  0.07}_{-  0.45-  0.07}$
&  4.40$^{+  0.50+  0.07}_{-  0.45-  0.07}$
&  6.39$^{+  0.59+  0.10}_{-  0.55-  0.10}$
&  5.77$^{+  0.58+  0.09}_{-  0.54-  0.09}$
&  6.79$^{+  0.58+  0.11}_{-  0.53-  0.11}$
&  9.39$^{+  0.97+  0.15}_{-  0.91-  0.15}$
&  10.3$^{+   1.0+   0.2}_{-   1.0-   0.2}$
 \\
    1.90--    2.20
&  1.35$^{+  0.27+  0.02}_{-  0.24-  0.02}$
&  2.11$^{+  0.32+  0.03}_{-  0.30-  0.03}$
&  2.31$^{+  0.24+  0.04}_{-  0.22-  0.04}$
&  2.85$^{+  0.30+  0.05}_{-  0.28-  0.05}$
&  2.71$^{+  0.36+  0.04}_{-  0.33-  0.04}$
&  3.17$^{+  0.39+  0.05}_{-  0.36-  0.05}$
&  3.28$^{+  0.39+  0.05}_{-  0.36-  0.05}$
&  4.16$^{+  0.43+  0.07}_{-  0.41-  0.07}$
&  4.72$^{+  0.47+  0.08}_{-  0.44-  0.08}$
&  5.22$^{+  0.46+  0.08}_{-  0.44-  0.08}$
&  7.96$^{+  0.82+  0.13}_{-  0.77-  0.13}$
&  8.31$^{+  0.84+  0.13}_{-  0.79-  0.13}$
 \\
    2.20--    2.55
&  1.36$^{+  0.24+  0.02}_{-  0.22-  0.02}$
&  1.38$^{+  0.24+  0.02}_{-  0.22-  0.02}$
&  2.04$^{+  0.20+  0.03}_{-  0.19-  0.03}$
&  2.19$^{+  0.25+  0.03}_{-  0.23-  0.03}$
&  2.05$^{+  0.30+  0.03}_{-  0.26-  0.03}$
&  2.45$^{+  0.31+  0.04}_{-  0.29-  0.04}$
&  2.47$^{+  0.31+  0.04}_{-  0.29-  0.04}$
&  2.72$^{+  0.33+  0.04}_{-  0.31-  0.04}$
&  3.58$^{+  0.38+  0.06}_{-  0.36-  0.06}$
&  3.48$^{+  0.35+  0.06}_{-  0.33-  0.06}$
&  4.33$^{+  0.57+  0.07}_{-  0.53-  0.07}$
&  5.83$^{+  0.65+  0.09}_{-  0.61-  0.09}$
 \\
    2.55--    2.96
&.768$^{+.178+.012}_{-.143-.012}$
&  1.12$^{+  0.21+  0.02}_{-  0.19-  0.02}$
&  1.46$^{+  0.16+  0.02}_{-  0.15-  0.02}$
&  1.69$^{+  0.20+  0.03}_{-  0.19-  0.03}$
&  1.32$^{+  0.23+  0.02}_{-  0.19-  0.02}$
&  2.46$^{+  0.30+  0.04}_{-  0.26-  0.04}$
&  2.12$^{+  0.27+  0.03}_{-  0.25-  0.03}$
&  3.00$^{+  0.32+  0.05}_{-  0.30-  0.05}$
&  2.72$^{+  0.31+  0.04}_{-  0.29-  0.04}$
&  2.97$^{+  0.30+  0.05}_{-  0.28-  0.05}$
&  3.65$^{+  0.49+  0.06}_{-  0.45-  0.06}$
&  4.42$^{+  0.53+  0.07}_{-  0.49-  0.07}$
 \\
    2.96--    3.44
&.538$^{+.130+.008}_{-.115-.008}$
&.698$^{+.140+.010}_{-.124-.010}$
&.990$^{+.118+.014}_{-.111-.014}$
&  1.04$^{+  0.14+  0.02}_{-  0.12-  0.02}$
&  1.57$^{+  0.21+  0.02}_{-  0.19-  0.02}$
&  1.19$^{+  0.19+  0.02}_{-  0.16-  0.02}$
&  1.52$^{+  0.20+  0.02}_{-  0.18-  0.02}$
&  1.63$^{+  0.22+  0.02}_{-  0.20-  0.02}$
&  2.04$^{+  0.25+  0.03}_{-  0.22-  0.03}$
&  2.04$^{+  0.21+  0.03}_{-  0.20-  0.03}$
&  2.73$^{+  0.37+  0.04}_{-  0.34-  0.04}$
&  3.97$^{+  0.44+  0.06}_{-  0.41-  0.06}$
 \\
    3.44--    3.99
&.567$^{+.122+.008}_{-.109-.008}$
&.765$^{+.141+.011}_{-.128-.011}$
&.685$^{+.091+.010}_{-.085-.010}$
&.906$^{+.123+.013}_{-.114-.013}$
&.622$^{+.125+.009}_{-.111-.009}$
&.998$^{+.152+.014}_{-.139-.014}$
&.910$^{+.151+.013}_{-.138-.013}$
&  1.28$^{+  0.17+  0.02}_{-  0.16-  0.02}$
&  1.34$^{+  0.18+  0.02}_{-  0.17-  0.02}$
&  1.57$^{+  0.18+  0.02}_{-  0.16-  0.02}$
&  1.89$^{+  0.30+  0.03}_{-  0.27-  0.03}$
&  2.21$^{+  0.32+  0.03}_{-  0.30-  0.03}$
 \\
    3.99--    4.63
&.361$^{+.096+.005}_{-.075-.005}$
&.391$^{+.094+.006}_{-.083-.006}$
&.544$^{+.074+.008}_{-.068-.008}$
&.612$^{+.094+.009}_{-.086-.009}$
&.590$^{+.116+.009}_{-.105-.009}$
&.697$^{+.122+.010}_{-.110-.010}$
&.587$^{+.112+.008}_{-.101-.008}$
&.901$^{+.141+.013}_{-.130-.013}$
&.856$^{+.136+.012}_{-.124-.012}$
&  1.14$^{+  0.14+  0.02}_{-  0.13-  0.02}$
&  1.24$^{+  0.22+  0.02}_{-  0.20-  0.02}$
&  1.83$^{+  0.26+  0.03}_{-  0.24-  0.03}$
 \\
    4.63--    5.38
&.186$^{+.067+.003}_{-.049-.003}$
&.138$^{+.058+.002}_{-.041-.002}$
&.246$^{+.048+.004}_{-.044-.004}$
&.506$^{+.081+.007}_{-.069-.007}$
&.397$^{+.092+.006}_{-.074-.006}$
&.523$^{+.097+.008}_{-.087-.008}$
&.521$^{+.097+.007}_{-.087-.007}$
&.605$^{+.106+.009}_{-.096-.009}$
&.801$^{+.125+.012}_{-.115-.012}$
&.634$^{+.097+.009}_{-.088-.009}$
&.813$^{+.163+.012}_{-.145-.012}$
&  1.15$^{+  0.20+  0.02}_{-  0.18-  0.02}$
 \\
    5.38--    6.24
&.174$^{+.058+.003}_{-.050-.003}$
&.356$^{+.077+.005}_{-.068-.005}$
&.251$^{+.045+.004}_{-.041-.004}$
&.356$^{+.061+.005}_{-.055-.005}$
&.301$^{+.073+.004}_{-.064-.004}$
&.279$^{+.070+.004}_{-.062-.004}$
&.277$^{+.070+.004}_{-.062-.004}$
&.414$^{+.084+.006}_{-.076-.006}$
&.565$^{+.094+.008}_{-.085-.008}$
&.626$^{+.089+.009}_{-.082-.009}$
&.724$^{+.152+.010}_{-.125-.010}$
&.840$^{+.156+.012}_{-.140-.012}$
 \\
    6.24--    7.25
&.092$^{+.038+.001}_{-.031-.001}$
&.193$^{+.055+.003}_{-.048-.003}$
&.171$^{+.036+.002}_{-.029-.002}$
&.215$^{+.045+.003}_{-.040-.003}$
&.224$^{+.057+.003}_{-.049-.003}$
&.251$^{+.061+.004}_{-.054-.004}$
&.273$^{+.060+.004}_{-.053-.004}$
&.311$^{+.067+.004}_{-.060-.004}$
&.333$^{+.069+.005}_{-.062-.005}$
&.412$^{+.072+.006}_{-.060-.006}$
&.520$^{+.121+.007}_{-.097-.007}$
&.405$^{+.103+.006}_{-.089-.006}$
 \\
    7.25--    8.41
&.089$^{+.038+.001}_{-.026-.001}$
&.049$^{+.027+.001}_{-.022-.001}$
&.133$^{+.029+.002}_{-.026-.002}$
&.152$^{+.035+.002}_{-.031-.002}$
&.183$^{+.049+.003}_{-.043-.003}$
&.187$^{+.048+.003}_{-.041-.003}$
&.167$^{+.047+.002}_{-.041-.002}$
&.248$^{+.058+.004}_{-.046-.004}$
&.277$^{+.060+.004}_{-.053-.004}$
&.271$^{+.053+.004}_{-.048-.004}$
&.448$^{+.104+.006}_{-.084-.006}$
&.404$^{+.097+.006}_{-.086-.006}$
 \\
    8.41--    9.76
&.043$^{+.024+.001}_{-.019-.001}$
&.093$^{+.032+.001}_{-.027-.001}$
&.102$^{+.023+.001}_{-.020-.001}$
&.177$^{+.036+.003}_{-.032-.003}$
&.097$^{+.033+.001}_{-.028-.001}$
&.076$^{+.032+.001}_{-.023-.001}$
&.093$^{+.032+.001}_{-.027-.001}$
&.086$^{+.033+.001}_{-.028-.001}$
&.230$^{+.051+.003}_{-.046-.003}$
&.175$^{+.039+.003}_{-.034-.003}$
&.232$^{+.067+.003}_{-.057-.003}$
&.301$^{+.076+.004}_{-.066-.004}$
 \\
      \hline
      \hline
    \end{tabular}
}
  \end{center}
}
\end{table}

\begin{table}[b!]
\tiny{
  \caption
{Observed positive muon fluxes}
  \label{tab:summp1}
\renewcommand{\arraystretch}{1.5}
  \begin{center}
\rotatebox{90}{
    \begin{tabular}{cllllllllllll}
      \hline
      \hline
      \begin{tabular}{@{}c@{}}       
      Momentum range\\
      (GeV/$c$)
      \end{tabular}
      & 
      \multicolumn{12}{c}{
      \begin{tabular}{@{}c@{}}       
	Flux$\pm~\Delta$Flux$_{\rm sta}
	\pm~\Delta$Flux$_{\rm sys}$\\
	(m$^{~-2}$sr$^{~-1}$s$^{~-1}$(GeV/$c$)$^{~-1}$)
      \end{tabular}
      }
      \\
      \hline
      &       \multicolumn{12}{c}{Atmospheric depth range [mean] (g/cm$^2$)}\\
      &  \multicolumn{1}{c}{4.46--4.80}  
      &  \multicolumn{1}{c}{4.86--7.21}  
      &  \multicolumn{1}{c}{7.04--8.23}  
      &  \multicolumn{1}{c}{8.24--9.08}  
      &  \multicolumn{1}{c}{9.06--9.54}  
      &  \multicolumn{1}{c}{9.60--11.4} 
      &  \multicolumn{1}{c}{11.4--12.5}
      &  \multicolumn{1}{c}{12.7--14.6}  
      &  \multicolumn{1}{c}{14.7--16.4} 
      &  \multicolumn{1}{c}{16.5--19.8}   
      &  \multicolumn{1}{c}{21.2--25.0}   
      &  \multicolumn{1}{c}{25.0--28.2}   \\
      & \multicolumn{1}{c}{[4.58]} & \multicolumn{1}{c}{[5.82]} 
      & \multicolumn{1}{c}{[7.81]} & \multicolumn{1}{c}{[8.70]} 
      & \multicolumn{1}{c}{[9.33]} & \multicolumn{1}{c}{[10.4]} 
      & \multicolumn{1}{c}{[11.9]} & \multicolumn{1}{c}{[13.6]} 
      & \multicolumn{1}{c}{[15.6]}& \multicolumn{1}{c}{[17.6]}
      & \multicolumn{1}{c}{[23.4]}& \multicolumn{1}{c}{[26.4]}\\
      \hline
    0.50--    0.58
&  13.2$^{+   4.0+   0.7}_{-   3.0-   0.7}$
&  15.9$^{+   4.0+   0.6}_{-   3.5-   0.6}$
&  11.4$^{+   3.3+   0.4}_{-   2.8-   0.4}$
&  22.6$^{+   3.8+   0.8}_{-   3.5-   0.8}$
&  20.0$^{+   4.6+   0.7}_{-   4.1-   0.7}$
&  28.4$^{+   5.3+   1.0}_{-   4.7-   1.0}$
&  25.3$^{+   5.0+   1.0}_{-   4.5-   1.0}$
&  35.1$^{+   6.1+   1.2}_{-   5.1-   1.2}$
&  31.6$^{+   5.5+   1.2}_{-   5.0-   1.2}$
&  37.0$^{+   5.5+   1.2}_{-   5.0-   1.2}$
&  51.9$^{+   9.4+   1.6}_{-   8.4-   1.6}$
&  62.2$^{+  10.3+   1.9}_{-   9.4-   1.9}$
 \\
    0.58--    0.67
&  9.84$^{+  2.83+  0.37}_{-  2.41-  0.37}$
&  10.5$^{+   3.1+   0.3}_{-   2.4-   0.3}$
&  15.4$^{+   3.4+   0.5}_{-   3.1-   0.5}$
&  24.4$^{+   3.5+   0.7}_{-   3.2-   0.7}$
&  18.5$^{+   4.0+   0.5}_{-   3.6-   0.5}$
&  21.9$^{+   4.2+   0.6}_{-   3.8-   0.6}$
&  31.0$^{+   4.9+   0.9}_{-   4.2-   0.9}$
&  25.1$^{+   4.5+   0.7}_{-   4.1-   0.7}$
&  26.0$^{+   4.5+   0.7}_{-   4.1-   0.7}$
&  28.0$^{+   4.4+   0.7}_{-   4.0-   0.7}$
&  47.5$^{+   8.1+   1.2}_{-   7.3-   1.2}$
&  35.2$^{+   7.1+   0.9}_{-   6.3-   0.9}$
 \\
    0.67--    0.78
&  9.08$^{+  2.56+  0.27}_{-  2.24-  0.27}$
&  11.6$^{+   2.8+   0.3}_{-   2.5-   0.3}$
&  18.6$^{+   3.3+   0.5}_{-   2.9-   0.5}$
&  20.7$^{+   3.0+   0.5}_{-   2.7-   0.5}$
&  21.0$^{+   3.8+   0.5}_{-   3.4-   0.5}$
&  23.5$^{+   3.9+   0.5}_{-   3.6-   0.5}$
&  21.7$^{+   3.8+   0.5}_{-   3.2-   0.5}$
&  22.4$^{+   3.9+   0.5}_{-   3.3-   0.5}$
&  27.6$^{+   4.3+   0.6}_{-   4.0-   0.6}$
&  26.4$^{+   3.8+   0.6}_{-   3.5-   0.6}$
&  38.6$^{+   6.6+   0.8}_{-   6.0-   0.8}$
&  46.4$^{+   7.2+   1.0}_{-   6.6-   1.0}$
 \\
    0.78--    0.90
&  4.40$^{+  1.67+  0.11}_{-  1.42-  0.11}$
&  12.2$^{+   2.5+   0.3}_{-   2.3-   0.3}$
&  8.58$^{+  2.07+  0.20}_{-  1.83-  0.20}$
&  17.5$^{+   2.4+   0.4}_{-   2.2-   0.4}$
&  14.5$^{+   2.9+   0.3}_{-   2.6-   0.3}$
&  15.8$^{+   3.0+   0.3}_{-   2.5-   0.3}$
&  17.2$^{+   2.9+   0.3}_{-   2.6-   0.3}$
&  23.2$^{+   3.4+   0.5}_{-   3.1-   0.5}$
&  25.1$^{+   3.6+   0.5}_{-   3.3-   0.5}$
&  33.0$^{+   3.7+   0.6}_{-   3.4-   0.6}$
&  42.1$^{+   6.2+   0.8}_{-   5.7-   0.8}$
&  45.5$^{+   6.8+   0.8}_{-   5.8-   0.8}$
 \\
    0.90--    1.05
&  4.03$^{+  0.72+  0.12}_{-  0.65-  0.12}$
&  8.05$^{+  0.99+  0.24}_{-  0.92-  0.24}$
&  10.2$^{+   0.8+   0.3}_{-   0.7-   0.3}$
&  12.2$^{+   1.0+   0.4}_{-   0.9-   0.4}$
&  11.2$^{+   1.2+   0.3}_{-   1.1-   0.3}$
&  15.4$^{+   1.4+   0.4}_{-   1.3-   0.4}$
&  16.3$^{+   1.4+   0.5}_{-   1.3-   0.5}$
&  17.3$^{+   1.5+   0.5}_{-   1.4-   0.5}$
&  18.1$^{+   1.5+   0.5}_{-   1.4-   0.5}$
&  21.2$^{+   1.5+   0.6}_{-   1.4-   0.6}$
&  28.1$^{+   2.5+   0.8}_{-   2.3-   0.8}$
&  31.4$^{+   2.7+   0.9}_{-   2.5-   0.9}$
 \\
    1.05--    1.21
&  6.07$^{+  0.75+  0.14}_{-  0.70-  0.14}$
&  7.98$^{+  0.88+  0.18}_{-  0.82-  0.18}$
&  9.18$^{+  0.67+  0.21}_{-  0.62-  0.21}$
&  10.3$^{+   0.8+   0.2}_{-   0.8-   0.2}$
&  11.1$^{+   1.0+   0.2}_{-   1.0-   0.2}$
&  11.7$^{+   1.0+   0.3}_{-   1.0-   0.3}$
&  14.1$^{+   1.2+   0.3}_{-   1.1-   0.3}$
&  16.4$^{+   1.3+   0.4}_{-   1.2-   0.4}$
&  16.5$^{+   1.3+   0.4}_{-   1.2-   0.4}$
&  16.7$^{+   1.2+   0.4}_{-   1.1-   0.4}$
&  23.3$^{+   2.1+   0.5}_{-   1.9-   0.5}$
&  26.9$^{+   2.2+   0.6}_{-   2.1-   0.6}$
 \\
    1.21--    1.41
&  5.28$^{+  0.66+  0.10}_{-  0.61-  0.10}$
&  5.02$^{+  0.65+  0.09}_{-  0.57-  0.09}$
&  7.82$^{+  0.56+  0.15}_{-  0.52-  0.15}$
&  8.15$^{+  0.67+  0.15}_{-  0.62-  0.15}$
&  9.16$^{+  0.85+  0.17}_{-  0.81-  0.17}$
&  10.7$^{+   0.9+   0.2}_{-   0.9-   0.2}$
&  10.7$^{+   0.9+   0.2}_{-   0.9-   0.2}$
&  13.4$^{+   1.0+   0.2}_{-   1.0-   0.2}$
&  13.0$^{+   1.0+   0.2}_{-   1.0-   0.2}$
&  15.6$^{+   1.0+   0.3}_{-   1.0-   0.3}$
&  20.3$^{+   1.7+   0.4}_{-   1.6-   0.4}$
&  20.8$^{+   1.8+   0.4}_{-   1.6-   0.4}$
 \\
    1.41--    1.63
&  3.12$^{+  0.46+  0.05}_{-  0.42-  0.05}$
&  3.89$^{+  0.51+  0.07}_{-  0.47-  0.07}$
&  6.23$^{+  0.46+  0.11}_{-  0.43-  0.11}$
&  6.95$^{+  0.56+  0.12}_{-  0.52-  0.12}$
&  6.98$^{+  0.68+  0.12}_{-  0.64-  0.12}$
&  7.34$^{+  0.69+  0.12}_{-  0.65-  0.12}$
&  9.03$^{+  0.78+  0.15}_{-  0.72-  0.15}$
&  9.91$^{+  0.82+  0.17}_{-  0.76-  0.17}$
&  11.7$^{+   0.9+   0.2}_{-   0.8-   0.2}$
&  10.9$^{+   0.8+   0.2}_{-   0.7-   0.2}$
&  15.9$^{+   1.4+   0.3}_{-   1.3-   0.3}$
&  17.9$^{+   1.5+   0.3}_{-   1.4-   0.3}$
 \\
    1.63--    1.90
&  3.27$^{+  0.44+  0.05}_{-  0.41-  0.05}$
&  3.22$^{+  0.43+  0.05}_{-  0.40-  0.05}$
&  4.51$^{+  0.36+  0.07}_{-  0.33-  0.07}$
&  5.72$^{+  0.47+  0.09}_{-  0.43-  0.09}$
&  5.62$^{+  0.57+  0.09}_{-  0.54-  0.09}$
&  5.60$^{+  0.56+  0.09}_{-  0.52-  0.09}$
&  7.56$^{+  0.66+  0.12}_{-  0.61-  0.12}$
&  7.61$^{+  0.66+  0.12}_{-  0.61-  0.12}$
&  8.67$^{+  0.72+  0.14}_{-  0.66-  0.14}$
&  9.60$^{+  0.68+  0.16}_{-  0.64-  0.16}$
&  12.1$^{+   1.1+   0.2}_{-   1.0-   0.2}$
&  12.1$^{+   1.1+   0.2}_{-   1.1-   0.2}$
 \\
    1.90--    2.20
&  2.18$^{+  0.33+  0.04}_{-  0.30-  0.04}$
&  2.77$^{+  0.36+  0.04}_{-  0.34-  0.04}$
&  4.13$^{+  0.31+  0.07}_{-  0.29-  0.07}$
&  3.45$^{+  0.33+  0.06}_{-  0.31-  0.06}$
&  4.93$^{+  0.49+  0.08}_{-  0.46-  0.08}$
&  4.28$^{+  0.45+  0.07}_{-  0.42-  0.07}$
&  5.50$^{+  0.51+  0.09}_{-  0.48-  0.09}$
&  5.69$^{+  0.52+  0.09}_{-  0.49-  0.09}$
&  6.22$^{+  0.55+  0.10}_{-  0.50-  0.10}$
&  7.21$^{+  0.54+  0.12}_{-  0.51-  0.12}$
&  8.92$^{+  0.86+  0.14}_{-  0.81-  0.14}$
&  11.3$^{+   1.0+   0.2}_{-   0.9-   0.2}$
 \\
    2.20--    2.55
&  1.63$^{+  0.25+  0.03}_{-  0.23-  0.03}$
&  1.93$^{+  0.29+  0.04}_{-  0.25-  0.04}$
&  3.44$^{+  0.26+  0.05}_{-  0.24-  0.05}$
&  2.85$^{+  0.28+  0.05}_{-  0.26-  0.05}$
&  3.12$^{+  0.35+  0.05}_{-  0.33-  0.05}$
&  3.57$^{+  0.38+  0.06}_{-  0.36-  0.06}$
&  4.63$^{+  0.43+  0.08}_{-  0.41-  0.08}$
&  3.92$^{+  0.40+  0.06}_{-  0.38-  0.06}$
&  5.03$^{+  0.44+  0.09}_{-  0.42-  0.09}$
&  5.57$^{+  0.44+  0.09}_{-  0.41-  0.09}$
&  6.97$^{+  0.72+  0.12}_{-  0.64-  0.12}$
&  7.95$^{+  0.77+  0.13}_{-  0.73-  0.13}$
 \\
      \hline
      \hline
    \end{tabular}
}
  \end{center}
}
\end{table}

\end{document}